\documentclass[preprint,5p,twocolumn]{elsarticle}

\usepackage{lineno,hyperref}
\usepackage{times}
\usepackage{epsfig}
\usepackage{graphicx}
\usepackage{amsmath}
\usepackage{amssymb}
\usepackage{diagbox}
\usepackage{caption}
\usepackage{mathrsfs}
\modulolinenumbers[5]

\begin{document}

\begin{frontmatter}

\title{End-to-end Learning for Joint Depth and Image Reconstruction from Diffracted Rotation}


\author[unipd]{Mazen Mel}
\cortext[mycorrespondingauthor]{Corresponding author}
\ead{mazen.mel@phd.unipd.it}
\author[sony]{ Muhammad Siddiqui}
\author[unipd]{Pietro Zanuttigh}

\address[unipd]{Department
of Information Engineering, University of Padova, Italy.}
\address[sony]{Sony Europe B.V., Zweigniederlassung Deutschland Stuttgart Technology Center, Stuttgart, Germany.}

\begin{abstract}
Monocular depth estimation is still an open challenge due to the ill-posed nature of the problem at hand. Deep learning based techniques have been extensively studied and proved capable of producing acceptable depth estimation accuracy even if the lack of meaningful and robust depth cues within single RGB input images severally limits their performance. Coded aperture-based methods using phase and amplitude masks encode strong depth cues within 2D images by means of depth-dependent Point Spread Functions (PSFs) at the price of a reduced image quality. In this paper, we propose a novel end-to-end learning approach for depth from diffracted rotation. A phase mask that produces a Rotating Point Spread Function (RPSF) as a function of defocus is jointly optimized with the weights of a depth estimation neural network. To this aim, we introduce a differentiable physical model of the aperture mask and exploit  an  accurate  simulation  of  the camera  imaging  pipeline. Our approach requires a significantly less complex model and less training data, yet it is superior to existing methods in the task of monocular depth estimation on indoor benchmarks. In addition, we address the problem of image degradation by incorporating a non-blind and non-uniform image deblurring module to recover the sharp all-in-focus image from its RPSF-blurred counterpart.
\end{abstract}


\end{frontmatter}

%\linenumbers

\section{Introduction}
Depth estimation from a single RGB image % captured by an on-board camera known as a passive monocular depth sensor. This 
is an ill-posed inverse problem, %mainly because images taken by a conventional camera convey very little 3D information  due to the nature of the image formation process. 
thus, additional image priors or sophisticated imaging systems are generally needed to account for the lack of depth cues in captured images. 
%One way to mitigate the limitations of conventional perspective cameras with regards to 3D information acquisition is to use more advanced computational imaging systems that are specifically engineered for the task of depth estimation. 
%In a nutshell, a computational camera is a more advanced version of its conventional counterpart, depending on the target task,
For example, the camera optics may be modified accordingly, a more advanced image sensor design and task-specific post processing algorithms may be developed in order to capture information well beyond the capabilities of conventional imaging systems. 
%Perhaps, among the well-known commercialized computational cameras today are
%An example are Light Field or Plenoptic cameras~\cite{adelson1992single,ng2005light} were an array of micro lenses placed in front of the image sensor captures information about light fields emanating from the scene allowing for interesting post capture image manipulation like refocus and variable aperture photography.

In this paper, we propose a computational camera  wherein the Point Spread Function (PSF) is altered in order to encode depth information within a 2D image. The desired PSF is obtained via an optimized phase mask inserted at the aperture plane of the camera. Our system encodes meaningful and robust depth cues in single RGB images thus making it easier for post-processing algorithms to produce accurate depth data. %, it requires a single RGB camera hence a small form factor and does not require a tedious configuration or calibration process, hence such system is suitable for a variety of robotics and mobile applications requiring energy constraints. 
However, such approach generally suffers from image quality degradation due to the poor light efficiency and/or the low Modulation Transfer Function (MTF) of the engineered PSF for high spatial frequency components.
In fact, amplitude aperture masks block a significant amount of light from reaching the image sensor resulting in low light throughput, while using phase only masks solves such problem since it only acts on the phase component of the incoming light wave. Still, the MTF drops rapidly in high frequency regions and the depth-dependent blurring caused by the camera's engineered PSF produces low SNR and degrade image quality.  

We propose a full pipeline that takes as input a single RGB image  and produces an estimated depth map of the scene along with the recovered sharp image from its PSF-blurred counterpart. We introduce a novel end-to-end deep learning model that jointly deals with the PSF design optimization and the depth estimation task. To this aim, a differentiable physical model of the aperture mask is  introduced together with an accurate simulation of the imaging pipeline including the optimized optics. In this way the learned model is able to firstly predict the optimal parameters of the phase mask design and then estimate the depth data from the RPSF-blurred input.  Finally, in order to address the image quality issue, a deep learning based non-blind and non-uniform deblurring module is incorporated.
% here we need to better introduce the work

%We present our approach for depth from diffracted rotation using end-to-end learning enabled by a differentiable physical model of the aperture mask in Section~\ref{sec:solution}.  %and~\ref{subsec:diff_mask_design}). 
%The Simulation process for generating image data is discussed in Section~\ref{sec:data} while the training procedures are presented in Section~\ref{sec:learning}. We show our simulation results for the tasks of monocular depth estimation and image restoration on synthetic data and compare our approach with the state-of-the-art %for the task of monocular depth estimation on real image data (
%in Section~\ref{sec:results} along with in-depth discussion of the model performance. At the end, an ablation study is presented in Section~\ref{sub_sec:ablation} to highlight the contributions of our method.

\section{Related Work}
\label{sec:related}
%------------------

% Should we include this ?

% this section the related work is discussed, firstly presenting the main works on the design of Rotating Point Spread Functions (RPSF) and then discussing methods to use their output for depth estimation. Finally we briefly overview methods for image reconstruction from the blurred output of RPSFs.

%------------------
\paragraph{Rotating Point Spread Functions}
\label{parag:1}
%------------------
RPSFs have been theoretically shown to increase the Fisher Information along the depth dimension by at least an order of magnitude as they have a uniformly lower Cramér–Rao bound across the axial dimension %compared to standard PSFs~
\cite{greengard2006depth}, which makes them highly sensitive to depth changes. RPSFs are obtained using pupil engineered cameras by means of phase and/or amplitude masks that are inspired by the concept of Orbital Angular Momentum (OAM) of light beams~\cite{allen1992orbital}. A beam with a rotating light intensity distribution along the propagation axis can be generated by a linear superposition of a set of suitably chosen Gaussian-Laguerre (GL) modes~\cite{piestun2000propagation,greengard2006depth} that can be optically encoded by the aperture mask. % which requires the pupil function to be modified in both its phase and amplitude components. 
These masks generate a PSF with invariant features that continuously rotate with defocus. However, such approach suffers from poor light throughput, some works  ~\cite{pavani2008high,shechtman2014optimal} addressed this problem and proposed iterative optimization schemes to ensure higher light efficiency of the engineered RPSF. Depth dependent RPSFs can be also generated by phase only masks.
For instance, a phase mask design inspired by Spiral Phase Plates (SPPs)~\cite{kotlyar2005generation,kotlyar2006diffraction} was introduced in~\cite{prasad2013rotating} where the pupil is subdivided into a set of annular Fresnel zones with an azimuthally increasing thickness profile, the delay imposed on the incident light waves increases azimuthally generating a corkscrew like wave-front carrying an OAM with a rotating phase function.
%of the form $e^{il\phi}$ where $l$ is the topological charge of the plate indicating how many spins the light does in a single wavelength, and $\phi$ is the azimuth angle in polar coordinates.
In~\cite{berlich2018high} and~\cite{kumar2015three}, the authors generalized the previous phase mask design by considering  a phase function that allows for generating multi-order-helix RPSFs by introducing new design parameters: the number of rotating lobes within the RPSF and the confinement of each zone, i.e. the inner and outer radii of each annular region, in addition to the number of Fresnel zones. While~\cite{berlich2018high} and~\cite{kumar2015three} used purely empirical approaches to determine the values for each design parameter, in this work we optimize those parameters in an end-to-end fashion jointly with the weights of a depth estimation deep neural network. 

%------------------
\paragraph{Depth estimation using coded apertures}
\label{parag:2}
%------------------
Levin et al.~\cite{levin2007image} proposed an amplitude modulation mask that was placed in front of a conventional camera lens to encode depth information via the diffracted pattern of the camera's PSF, %, which is basically the shadow of the mask itself, 
as point-like sources move along a plane parallel to that of the camera sensor, the mask shadow would shift accordingly and as they move closer or farther from the camera, the pattern would expand or shrink. This information is later used to determine the object's distance from the camera. The mask introduced by~\cite{levin2007image} has opaque regions blocking a significant amount of light from reaching the image sensor. Zhou et al.~\cite{zhou2011coded} built upon the work of Levin et al.~\cite{levin2007image} and introduced two complementary amplitude masks. % that were used to capture two different coded images that were combined to produce a full depth map.

In~\cite{greengard2006depth} the concept of depth from diffracted rotation was introduced: a superposition of a set of suitably chosen Gaussian-Laguerre modes generates a double helix RPSF that was used to estimate the depth of a planar scene by analyzing the blurring effects within the captured image. However, low MTF by the mask leads to poor SNR within the captured image, thus limiting the capability of signal-processing based algorithms to recover  sharp images or  accurate depth maps.
In addition, earlier studies relied on a design paradigm based on the separate optimization of the camera optics and  post-processing algorithms:  they design the mask first and then tailor a reconstruction algorithm that fits the proposed physical design of the mask as in~\cite{kumar2013psf,kumar2015three,roider2014axial,berlich2016single}. Such design methodology, however, leads to sub-optimal performance. 

%--------------------
\paragraph{End-to-end learning for monocular depth estimation}
%--------------------

Recently, an emerging trend appeared to tackle the separate design issue and new frameworks for joint optical and digital optimization using deep learning techniques have been introduced. These methods exploit end-to-end learning to optimize the mask’s height map  together with the trainable weights of a Convolutional Neural Network (CNN). Haim et al.~\cite{haim2018depth} proposed a differentiable phase mask consisting of concentric rings that introduce depth-dependent chromatic aberrations and encoding depth cues within single captured images. In the work of Chang et al.~\cite{chang2019deep} and Wu et al.~\cite{wu2019phasecam3d} a free-form differentiable phase mask design parameterized using a set of superposed Zernike polynomials~\cite{iskander2001optimal,born2013principles} was jointly optimized with the weights of a  U-Net~\cite{ronneberger2015u}. However, the employed camera model was not realistic accounting only for additive Gaussian noise~\cite{wu2019phasecam3d}. Furthermore, unrestricted parameterized mask design and higher degrees of freedom may cause the optimization to converge towards local minimas as the objective function becomes too complex leading to %generic PSF shapes with no clear correlations with different depth planes which may lead to
sub-optimal performance. In this work, the mask is parameterized using only three design parameters two of which are optimized in an end-to-end fashion, and the PSF has clear and simple correlations with depth.

%------------------
\paragraph{Image deblurring}
%------------------
%RPSF simulated images are blurred with a spatially  variant and depth-dependent blur kernel resulting in image quality degradation and undesirable blurring artifacts. In order to address this issue, a non-blind and non-uniform image deblurring method must be implemented. 

Non-blind image deblurring has been extensively studied before (e.g.,~\cite{anger2018modeling,cho2009fast,gong2020learning,krishnan2009fast}), and a substantial performance gain has been made easier using deep learning based image deblurring techniques.
%In order to recover the sharp image, a traditional approach would be to use a Wiener deconvolution filter~\cite{wiener1950extrapolation} which minimizes a regularized least square error between the recovered image and the desired output. Another possible solution is to use Richardson-Lucy~\cite{richardson1972bayesian} iterative algorithm to estimate the probability distribution of the sharp latent image from the blurry and noisy distribution using Bayes' theorem. Half-Quadratic Splitting (HQS) methods~\cite{geman1995nonlinear} can be used for non-blind image deblurring wherein an objective function containing a separate data fitting and an auxiliary regularization terms are jointly optimized in an iterative way.  

%Recently, deep learning based techniques have been extensively studied for non-blind image deblurring. 
%Provided a large number of training samples, a neural network is able to learn meaningful image priors in latent space, enabling accurate and faster image reconstruction. 
The authors of~\cite{xu2014deep} used the separability property of the pseudo-inverse kernel in Wiener deconvolution filter to design a dedicated CNN. %which performs 1D separable convolutions on input images.
In~\cite{schuler2013machine}, a two-stage deblurring process was introduced using a Wiener deconvolution filter %is applied to the input image and then
and a simple MLP architecture. % was trained to remove deconvolution artifacts. 
A novel deconvolution approach was recently introduced by~\cite{dong2021deep} where a Wiener deconvolution filter is applied to the input data in feature space. % ., i.e. on feature maps extracted from the original blurred image, and then a multi-scale feature refinement stage was implemented. % to produce the full resolution deblurred image from the deconvolved input features. 

\section{Depth Estimation from Diffracted Rotation}
\label{sec:solution}

%------------------ Full pipeline figure --------------------------
\begin{figure*}[h!]
\begin{center}
\includegraphics[width=0.8\linewidth]{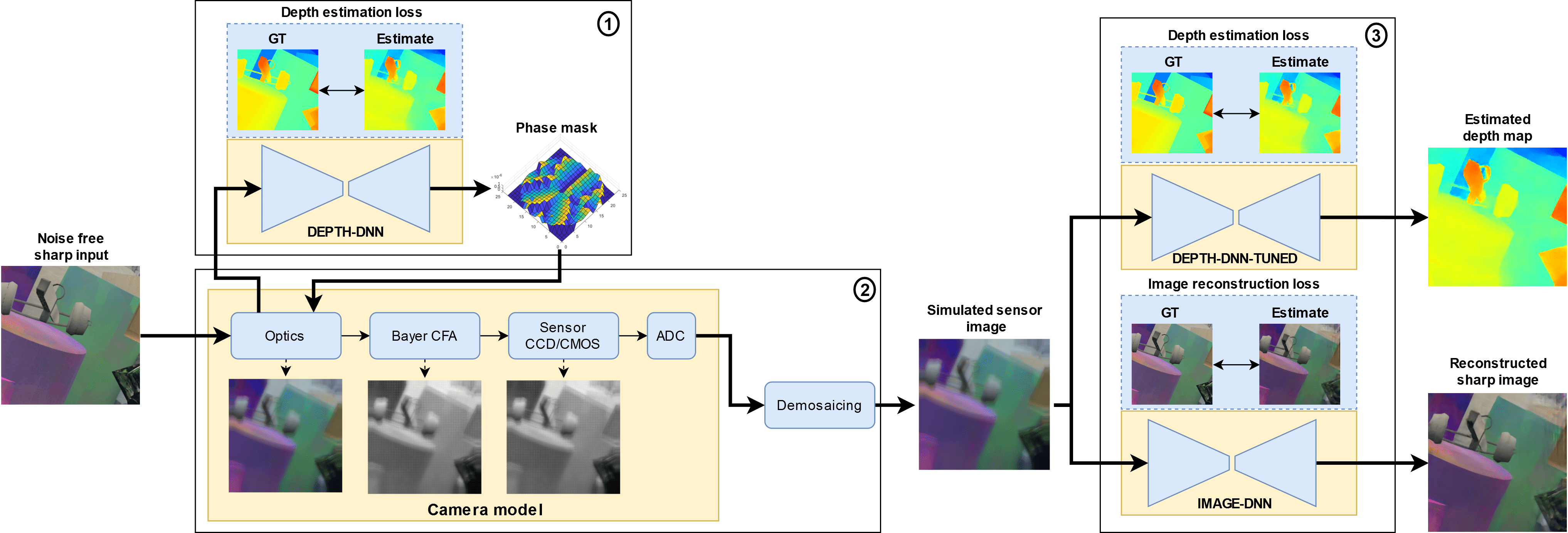}
\end{center}
   \caption{The full architecture of our end-to-end learning framework.}
\label{fig:fullpipeline}
\end{figure*}
%------------------------------------------------------------------

We propose an end-to-end learning approach for depth from diffracted rotation using RPSF-coded images. As shown in Fig.~\ref{fig:fullpipeline}, the full pipeline of the proposed solution encompasses three main stages. In the first stage, the height map of the phase mask is jointly learned  with the weights of a neural network (DEPTH-DNN) trained to perform monocular depth estimation. This module is trained using noise-free RPSF-blurred synthetic images. In the second stage, the optimized phase mask is fitted within the optics module and a digital image formation pipeline is applied to the RPSF-coded synthetic images in order to simulate a realistic camera model. Finally, we used the demosaiced images from the previous stage as input to fine-tune the weights of DEPTH-DNN on noisy data obtaining a refined model (DEPTH-DNN-TUNED) and to recover the all-in-focus sharp image using a dedicated network (IMAGE-DNN) which performs non-blind and non-uniform image deblurring. Both of these modules make up the third and last stage of the proposed architecture. 

%------------------
\subsection{Engineered PSF}
%------------------
In this section we briefly describe the effect of modified optics on the system's PSF. For more details please refer to the supplementary materials. 

For simplicity, consider an on-axis ideal point source situated at optical infinity, the light field $U_{in}$ with a constant amplitude $P$ and a phase $\psi$ emanating from such source has the form:
$U_{in}=Pe^{i\psi}$.
An optical element such as a lens or a phase mask with a refractive index $n$ and a height profile $h$ introduces a phase delay $\Phi = \frac{2\pi (n-1)}{\lambda}h$ on incident light wave-fronts. If a phase mask is inserted at the entrance pupil plane of an imaging system, the total phase delay can be expressed as the sum of the delay due to the lens with the one due to the mask: 
\begin{equation}
    \Phi_{optics} = \Phi_{lens} + \Phi_{mask}
\end{equation}

The  light field after the lens and mask system has the form:
\begin{equation}
    U_{out}=A\cdot P \cdot e^{i(\psi+\Phi_{optics})}
\end{equation}
Where $A$ is the aperture mask simulating the finite aperture area. The field $U_{out}$ can be further propagated to the image plane and the PSF can be obtained by the field's intensity distribution. 
Choosing the appropriate height profile of the phase mask helps design specific PSF patterns depending on the target task.
%-----------------------------
\subsection{Phase Mask Design}
\label{subsec:mask_design}
%-----------------------------

Both annular and free-form mask designs have been studied in the context of joint optimization of camera optics with post-processing algorithms~\cite{haim2018depth,wu2019phasecam3d,chang2019deep}. We build upon the depth-dependent RPSF introduced by~\cite{prasad2013rotating}. In this section, the mathematical model of the mask's phase profile is described and in the next section a differentiable approximation of the mask's height map is derived to allow for gradient back-propagation in our end-to-end learning framework.

We use a Fresnel-zone based design~\cite{prasad2013rotating} that has an outermost radius $R$ and $L$ phase plates in the form of concentric annular regions each of topological charge $l=1,...,L$ and bounded by two radii $R_{l-1}=R\sqrt{\frac{l-1}{L}}$ and $R_{l}=R\sqrt{\frac{l}{L}}$.

The pupil plane phase can be written as in \cite{prasad2013rotating}:

\begin{equation}
\label{4}
    \Phi_{mask}(\rho,\phi) = \begin{cases}
     \phi & 0 \leq \rho < \sqrt{\frac{1}{L}}\\
      & \vdots \\
     l\phi & \sqrt{\frac{l-1}{L}} \leq \rho < \sqrt{\frac{l}{L}}\\
      & \vdots \\
     L\phi & \sqrt{\frac{L-1}{L}} \leq \rho < 1
    \end{cases}
\end{equation}

$\Phi_{mask}$ is defined with the polar position vector $\rho$ normalized by the pupil's outermost radius $R$, and is a step function modeling the physical design property of concentric rings each  with its own phase profile. 

The phase profile in Eq.~\ref{4} can be further generalized to account for multi-order-helix RPSFs as in~\cite{kumar2015three} and~\cite{berlich2018high} by expanding it into:

\begin{equation}
\label{6}
    \Phi_{mask}(\rho,\phi) = \begin{cases}
     \phi & 0 \leq \rho < (\frac{1}{L})^{\epsilon}\\
      & \vdots \\
     [(l-1)N+1]\phi & (\frac{l-1}{L})^{\epsilon} \leq \rho < (\frac{l}{L})^{\epsilon}\\
      & \vdots \\
     [(L-1)N+1]\phi & (\frac{L-1}{L})^{\epsilon} \leq \rho < 1
    \end{cases}
\end{equation}

Notice that the inner and outermost radii of each zone are now controlled by $\epsilon$ which lies in $[0,1]$, and the topological charge of each ring is now $[(l-1)N+1]$ instead of just $l$. Besides the number of rings $L$, $N$ and $\epsilon$ are two new design parameters each having an effect over the resulting RPSF shape. More precisely, $N$ defines the number of peaks or the main rotating lobes of the RPSF and $L$ and $\epsilon$ both control the peak separation and confinement of each peak. %Fig.~\ref{fig:N_eps_L} (a) shows a set of multi-order-helix RPSFs at the same depth plane with different N values, 
In the case of a single helix rotating PSF $(N=1)$  the phase profile of each ring would be reduced to the original expression as in Eq.~\ref{4}. Notice also that by increasing the number of peaks, the practical depth range would be reduced because of rotation ambiguities when peaks rotate beyond $[-\frac{\pi}{N},\frac{\pi}{N}]$.

%------------------
\subsection{Differentiable Phase Mask Design}
\label{subsec:diff_mask_design}
%------------------
The phase profile presented in Eq.~\ref{6} is not differentiable with respect to the design parameters $N$ and $\epsilon$. Thus, a differentiable approximation for this equation is necessary in order to simulate the camera's optical layer and enable both forward and backward propagation. The number of Fresnel zones $L$ is considered as a hyper-parameter that can be manually tuned to achieve better depth estimation performance.

\begin{figure}[h!]
\centering
\includegraphics[width=\linewidth]{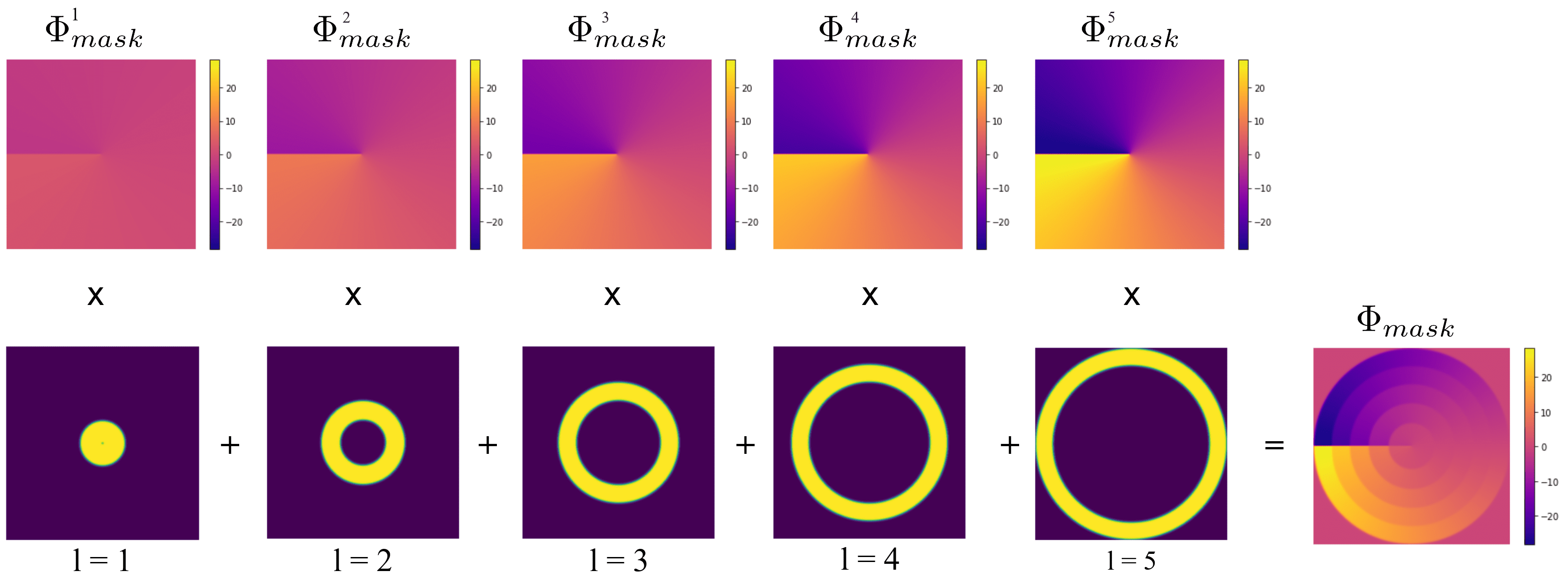}
\caption{A differentiable phase mask design where each Fresnel zone is simulated by a ring mask multiplied by the phase profile ($\Phi_{mask}^{i}$) corresponding to each zone.}
\label{fig:masks}
\end{figure}

The steps in Eq.~\ref{6} can be approximated with a set of 2D rings in polar coordinates with increased radii as $L$ increases as illustrated in Fig.~\ref{fig:masks}, each Fresnel zone is obtained by subtracting the areas of two 2D $\tanh$ functions each with a radius corresponding to the one of the two radiis of the desired ring. We multiply the inner coordinates of each $\tanh$ by a large constant (100 in our case) in order to get sharp mask edges. The new approximated phase profile equation for $L$ zones can be written as:

\begin{equation}
\begin{aligned}[b]
    & \tilde{\Phi}_{mask}(\rho,\phi)= \sum_{l=0}^{l=L}\underbrace{\bigr((l-1)N+1\bigl) \phi}_\text{$l^{th}$ ring phase} \times {}\\ &\underbrace{\frac{1}{2}\bigr(\tanh[100(\rho-r_{l})]-\tanh[100(\rho-r_{l+1})]\bigl)}_\text{$l^{th}$ ring mask}
    \end{aligned}
\end{equation}

Where $r_{0}=0$, $r_{l}=R(\frac{l}{L})^\epsilon ; ~\forall l\in[1..L]$, and $R$ is the outermost radius of the mask. $\phi$ and $\rho$ are the polar coordinates.

\begin{figure}[h!]
    \centering
    \includegraphics[width=\linewidth]{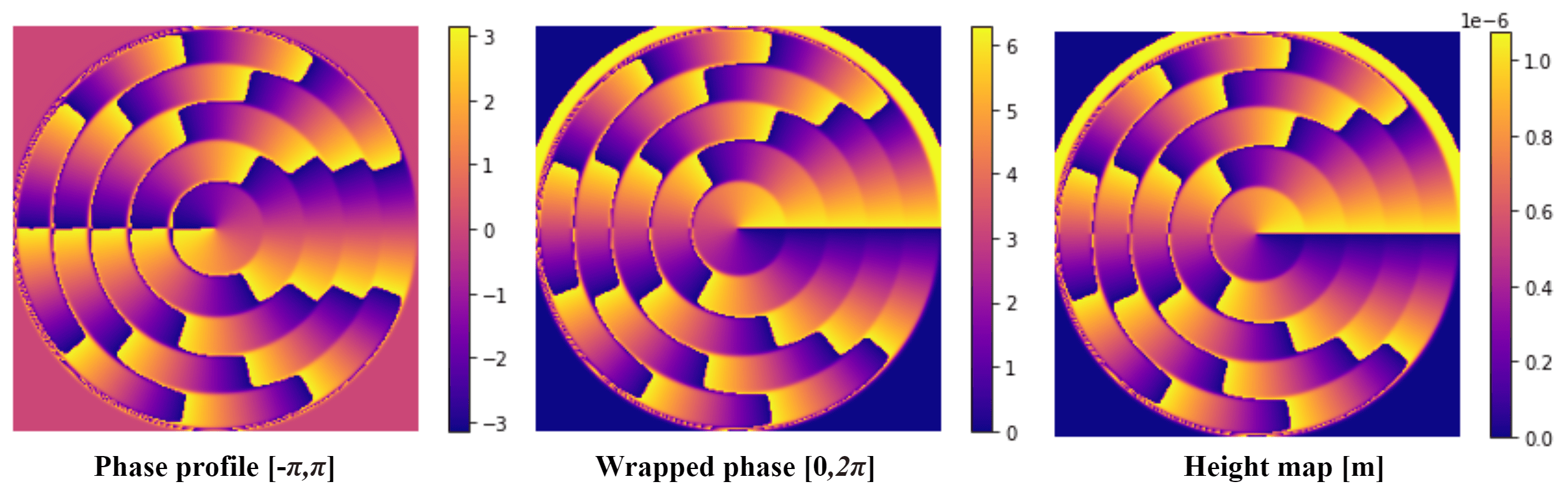}
    \caption{The height map of the phase mask. From left to right, the phase distribution as the argument of $e^{i\Tilde{\psi}}$, the $[0,2\pi]$ wrapped phase distribution, and the obtained height map for $[N=2,L=5,\epsilon=0.9]$.}
    \label{fig:height}
\end{figure}

Each phase profile $\Phi_{mask}^{l}=\bigr((l-1)N+1\bigl)\phi$ is multiplied by the corresponding ring mask and the resulting zones are added up to produce the final phase profile of the mask. We produce the height map $h$ of the mask as follows:

\begin{equation}
    h(x,y)=\frac{\lambda}{2\pi(n-1)}\cdot  \bigr(arg\{e^{i\tilde{\Phi}_{mask}}\} \mod 2\pi\bigl)
\end{equation}

where $arg$ is the complex argument function, and the modulo accounts for the phase wrapping operation (phase values have a $2\pi$ periodicity as shown in  Fig.~\ref{fig:height}).

Notice that the height profile of the phase mask is wavelength dependent: the resulting PSFs for the three RGB color primaries have different rotation rates which introduces chromatic aberrations. Still, it will not be problematic in the context of an end-to-end optimization framework since the network could learn the correlations between the rotation rate and the corresponding color channel. In fact, such aberrations can be seen as depth-dependent and can also relay valuable depth cues. For the physical design of the mask a reference wavelength value ($\lambda= 536.67 nm$) is used to produce the height map. 

\begin{figure}[h!]
    \centering
    \includegraphics[width=\linewidth]{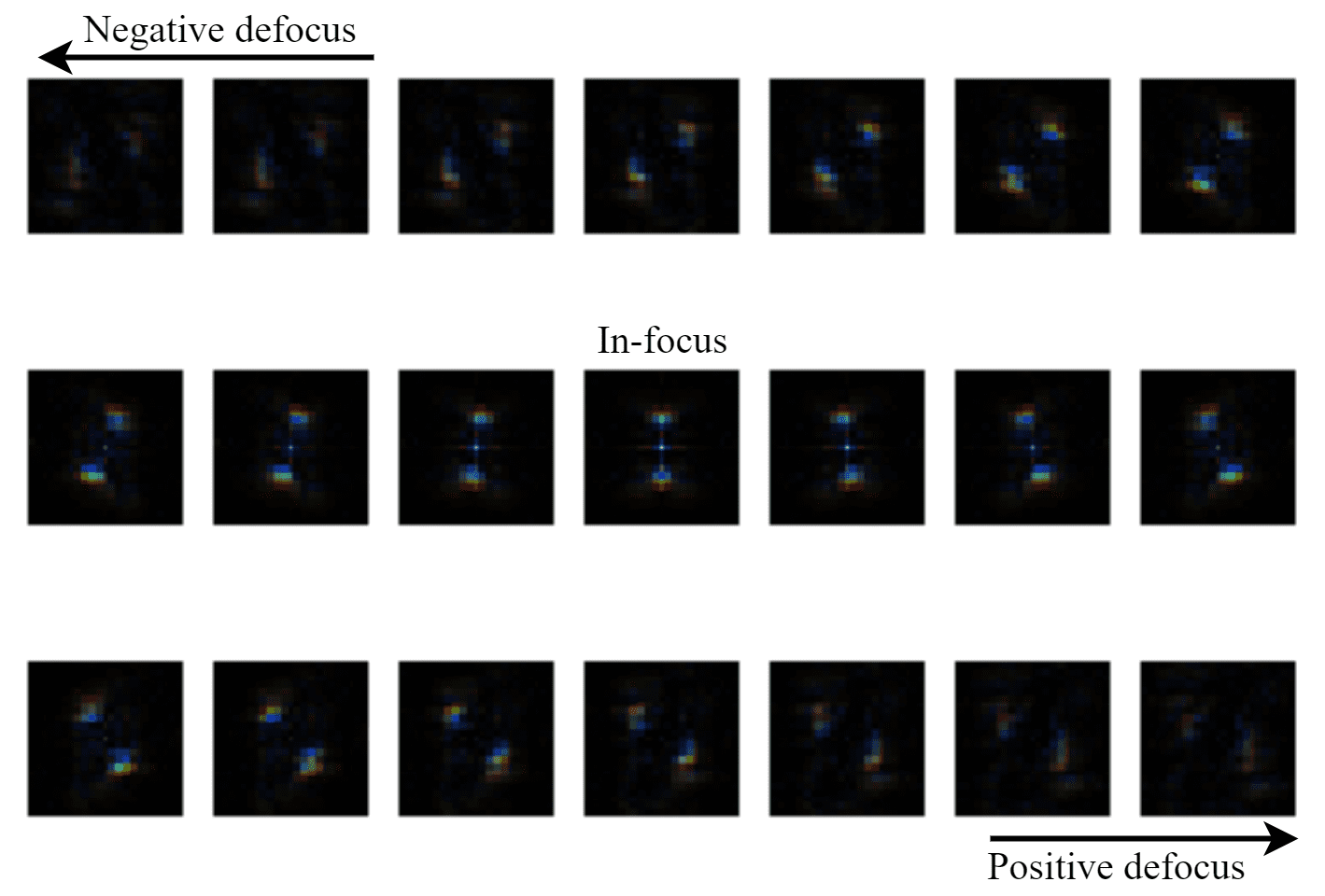}
    \caption{A double helix RPSF obtained by a mask with $[N=2,L=5,\epsilon=0.9]$. The RPSF's intensity distributions are shown as a function of defocus.}
    \label{fig:dhrpsf}
\end{figure}

Fig.~\ref{fig:dhrpsf} shows a double helix RPSF generated by a phase mask with $[N=2,L=5,\epsilon=0.9]$, it has two main lobes rotating counterclockwise as a function of defocus. Notice that even at the in-focus plane the RPSF has the same shape thus objects that are "in-focus" are also blurred.

\section{Training Data Generation}
\label{sec:data}
\subsection{Datasets}
A subset of FlyingThings3D dataset is used for the joint optimization of the phase mask and depth estimation network, and to train IMAGE-DNN. This dataset is a part of the Scene Flow synthetic datasets~\cite{MIFDB16}. This subset was previously used by~\cite{wu2019phasecam3d} in a similar joint optimization approach, it contains synthetic images of randomly placed objects with pixel-accurate disparity maps. The training, validation, and test sets contain respectively 5078, 555, and 420 images with a resolution of $278\times278$ pixels. Additionally, in order to evaluate the performance of our approach with the state-of-the-art in the task of monocular depth estimation on real data, NYUV2~\cite{eigen2014depth} depth dataset is used to train and evaluate DEPTH-DNN. Originally, the dataset contains 120k training images of indoor scenes with a resolution of $640\times480$ pixels along with ground truth depth maps acquired by a Microsoft Kinect V1 depth sensor. The test set as defined by the  split in~\cite{eigen2015predicting} contains 654 images. In this work, a subset of 50k samples of NYUV2 is used to train the depth estimation network as in~\cite{bhat2020adabins}. Finally, the test set of SUNRGBD dataset~\cite{song2015sun} is used to evaluate the generalization capability of DEPTH-DNN. This set contains 5050 test images of indoor scenes with ground truth depth maps acquired by four different depth sensors some of which use active illumination techniques and others incorporate passive stereo systems.

%------------------
\subsection{Image Formation Model}
\label{sec:image_formation}
%------------------
Light rays emanating  from the scene are acquired by the camera and optically coded by the phase mask via a depth-dependent blurring process with the camera's RPSF. For our simulations, the ground truth depth maps are approximated with a layered model in which only a finite number of depth planes are used to compose and render the RPSF-coded image using the following image formation model:

\begin{equation}
\label{18}
    I_{sim}=\sum_{d=1}^{d=D}(I_{aif}*RPSF_{d})\odot M_{d}
\end{equation}

Where $I_{sim}$ is the simulated blurred image, $I_{aif}$ is its all-in-focus counterpart, $RPSF_{d}$ is the RPSF intensity distribution at the depth plane $d$, $\odot$ stands for element-wise multiplication, and $\{M_{d};~d=1,...,D\}$ are the depth masks presenting the individual depth layers such that at each pixel location $\sum_{d}M_{d}=1$, i.e. only one pixel mask is set to one at each position. 
%better explain how many planes at which position, sum=1 i.e. 0001000 or avg ?? 

Afterwards, the image formation pipeline is applied to the RPSF-blurred images to simulate real cameras. 
As illustrated by Fig.~\ref{fig:fullpipeline}, a Bayer CFA receives the full color resolution RPSF-blurred image and produces down-sampled RGGB color pattern. Although it is hard to accurately simulate the noise behaviour within the sensor chip, the final amount of noise is mainly caused by sensor shot and read noises. To this end, we simulated the read noise with an additive Gaussian $\mathcal{N}(0,\,\sigma^{2})$ with zero mean and a fixed standard deviation $\sigma=0.01$, photon shot noise follows a Poisson distribution, in practice it is modeled by a Gaussian distribution whose mean and variance depend on the expected photon count over the exposure time. The resulting noisy sensor image is quantized with an ADC module with a resolution of 8 bits. Finally, a linear interpolation-based demosaicing technique of~\cite{malvar2004high} is used to recover the full color channels from the CFA and produce the final output which will be used to fine-tune DEPTH-DNN and train IMAGE-DNN.

%------------------
\section{Deep Learning Framework}
\label{sec:learning}
%-----------------
 
\subsection{Monocular Depth Estimation}

In the first stage of the proposed solution (see Fig.~\ref{fig:fullpipeline}), the phase mask design parameters $[N,\epsilon]$ are jointly learned with the weights of a U-Net~\cite{ronneberger2015u} which is trained on a subset of FlyingThings3D~\cite{MIFDB16}. Two different learning rates, $L_r^{mask}=0.1$ and $L_r^{dnn}=1e-4$ are set for the phase mask and for the depth estimation neural network. During the training process, the gradient error is back-propagated through the network as well as the mask's trainable parameters and the weights are updated accordingly using TensorFlow's automatic differentiation framework~\cite{tensorflow2015-whitepaper}. The network of this stage is trained for 150k iterations using a batch size of 20. Adam optimizer~\cite{kingma2014adam} is used with exponential decay rates of the first momentum and second momentum respectively set to $\beta_{1}=0.99$ and $\beta_{2}=0.999$. 

Similar to Wu et al.~\cite{wu2019phasecam3d}, we used a combination of Root Mean Square Error (RMSE) loss $\mathcal{L}_{rmse}$ and gradient loss $\mathcal{L}_{grad}$ which forces the network to estimate accurate depth maps with well defined object boundaries at different depth planes. The total loss function is defined as:

\begin{equation}
    \mathcal{L}_{depth} = \mathcal{L}_{rmse} + \mathcal{L}_{grad}
\end{equation}

Where $\mathcal{L}_{rmse}$ and $\mathcal{L}_{grad}$ are defined as:

\begin{equation}
    \mathcal{L}_{rmse}(\theta,\theta^{*}) = \sqrt{\frac{1}{|T|}\sum_{\theta\in T}(\theta-\theta^{*})^{2}}
\end{equation}

\begin{equation}
    \mathcal{L}_{grad} (\theta,\theta^{*}) =\mathcal{L}_{rmse}(\frac{\partial \theta}{\partial x},\frac{\partial \theta^{*}}{\partial x})+\mathcal{L}_{rmse}(\frac{\partial \theta}{\partial y},\frac{\partial \theta^{*}}{\partial y})
\end{equation}

$\theta$ and $\theta^{*}$ are respectively the predicted and the ground truth disparity maps and the subtraction is done pixel-by-pixel, $x$ and $y$ are the spatial dimensions, and $|T|$ is the number of disparity maps.

In the third stage, the same U-Net~\cite{ronneberger2015u} which was previously trained on noise-free RPSF data is fine-tuned using 50k training iterations with noisy images simulated by the camera model of the second stage (see Fig.~\ref{fig:fullpipeline}), the phase mask is fixed during this training pass and only the network's weights are updated. All hyper-parameters' values are fixed to the same values as in the first training stage.

For NYUV2 dataset~\cite{eigen2014depth}, both the network and the mask are learned  using 50k training samples, the input images to the network and the output depth maps have a resolution of $320\times240$ which correspond to half of that of the original samples, a bilinear-upsampling is applied to the predicted depth maps to recover the original resolution of $640\times480$ for evaluation purposes as in \cite{bhat2020adabins, chen2019structure,hao2018detail}. The network is trained for 150k iterations with a batch size of 20, the phase mask and  the neural network learning rates as well as the optimizer used in the training are the same as the ones used for the subset of FlyingThings3D~\cite{MIFDB16}.

%------------------
\subsection{Image Deblurring}
\label{subsec:image_deblurring}
%------------------

We used a modified version of the deblurring model proposed by~\cite{dong2021deep} wherein a Wiener deconvolution is applied to a set of feature maps extracted from the blurred input using a simple CNN architecture as feature extractor, the deconvolved feature maps are then fed into a multi-scale feature refinement stage in order to get the final deblurred image in a coarse-to-fine based reconstruction technique. Such approach proved capable of restoring very fine structural details allowing for accurate image reconstruction. In this work, we adapted the network proposed by~\cite{dong2021deep} for the case of non-uniform image deblurring as the blur kernel in our case is spatially variant. More precisely, a separate deconvolution is performed for every depth layer and the results are cropped using the corresponding depth masks $M_{d}$ as in Eq.~\ref{18}. The deconvolved feature maps are then combined in order to get the final Wiener filter output. We found that, even though the Wiener filter module is applied in feature space, some undesirable deconvolution artifacts, most noticeably ringings, are visible especially around image boundaries and object edges. Since the $FFT$ operator within the Wiener deconvolution filter supposes circular periodicity of the input. To tackle this problem, an edgetaper operation~\cite{reeves2005fast} was implemented on the blurred input image to smooth out its boundaries which can considerably reduce the ringing artifacts in the final reconstructed sharp image. 

We trained the network using the subset of FlyingThings3D~\cite{MIFDB16} for 500 epochs with Adam optimizer~\cite{kingma2014adam} with exponential decay rates of the first momentum and second momentum respectively set to $\beta_{1}=0.99$ and $\beta_{2}=0.999$, and a learning rate $L_r^{image}= 1e-4$ which is halved after 250 epochs. The number of auto-encoders in the multi-scale feature refinement modules is set to 2 as in the original work of~\cite{dong2021deep},  the number of extracted feature maps from the blurry input is 16, and the batch size is set to 8. For the loss function, it was experimentally seen that $L{1}$ norm leads to better reconstruction results than the ones obtained with $L{2}$ norm.  

\begin{equation}
    \mathcal{L}_{image} (\theta,\theta^{*}) =\frac{1}{|T|}\sum_{\theta\in T}|\theta-\theta^{*}|
\end{equation}

Where $\theta$ and $\theta^{*}$ are respectively the reconstructed and the ground truth sharp images and $|T|$ is the number of images.

\section{Experimental Results}
\label{sec:results}

%This section  presents the experimental evaluation of the proposed approach on synthetic data for the tasks of monocular depth estimation and non-blind image deblurring. A comparison with the state-of-the-art is reported on two real indoors datasets for the task of monocular depth estimation. Finally, an ablation study is presented to showcase the performance gains due to the various components of the proposed approach. %that could be achieved using RPSFs for depth estimation.  %. %Depth estimation results are shown 

%This section  presents the experimental evaluation of the proposed approach. Depth estimation results on three different datasets are shown followed by a brief evaluation of the color image restoration module and by the ablation study.

%------------------
\subsection{Experimental Results on Synthetic Data}
\label{subsec:depth_results}
%----------------------------------

\begin{figure}[h!]
\centering
\includegraphics[width=\linewidth]{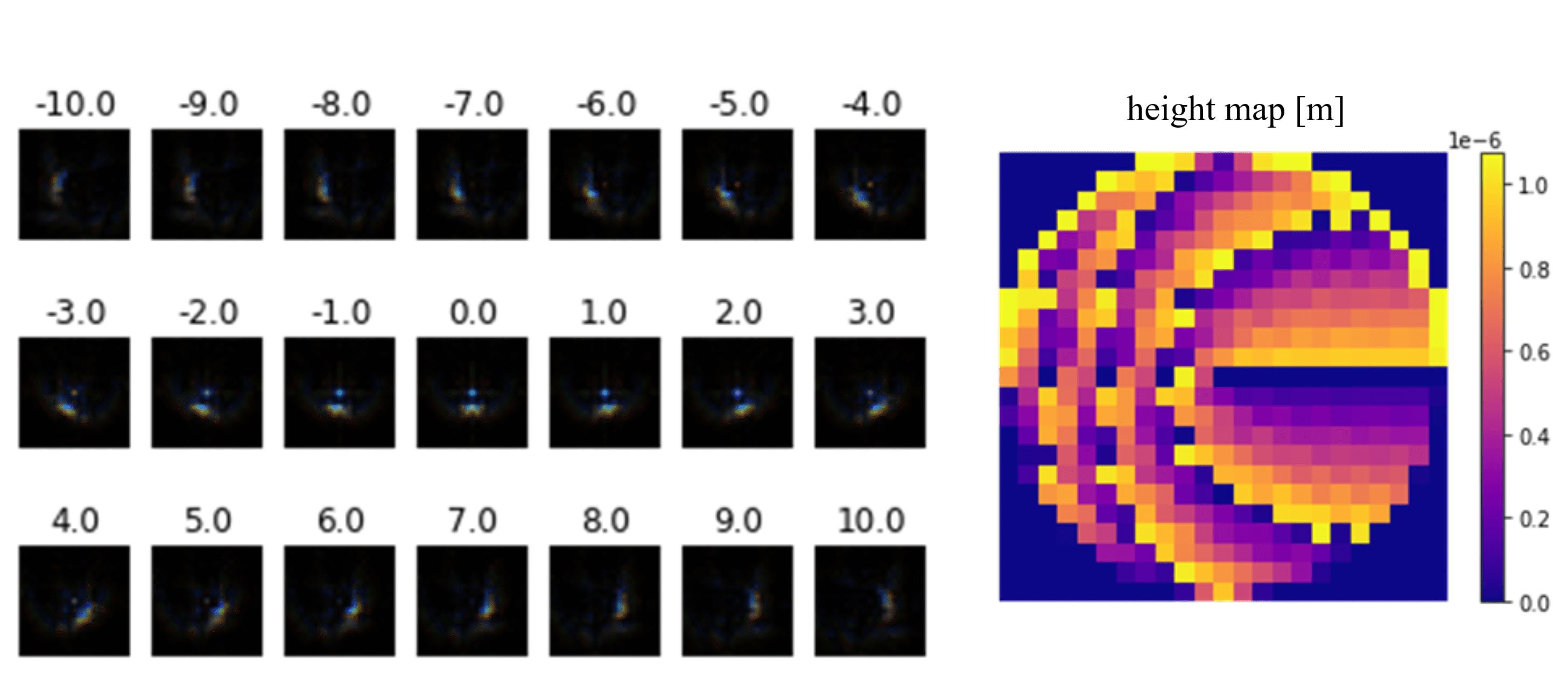}
\caption{Learned RPSF shape at different defocus planes (left),  height map of the phase mask (right).}
\label{fig:learndedrpsf1}
\end{figure}

%----------------------------------
\paragraph{Monocular depth estimation on FlyingThings3D subset}
%----------------------------------
%The quantitative results for this dataset are reported in Table.~\ref{tab:flayingthings1} while qualitative results samples are shown in Fig.~\ref{fig:res1} for the case of noise free as well as noisy inputs. 
%\begin{table}[h!]
%\begin{center}
%\begin{tabular}{|l||c|}
%\hline
% Input & RMSE~$\downarrow$ \\
%\hline\hline
%Noise-free images & 0.392\\
%Noisy images & 0.712\\
%\hline
%\end{tabular}
%\end{center}
%\caption{Quantitative results of monocular depth estimation on the test set of the subset of FlyingThings3D~\cite{MIFDB16}.}
%\label{tab:flayingthings1}
%\end{table}
%\textit{Monocular depth estimation on FlyingThings3D subset:}\\
For this set of experiments the phase mask's trainable parameters are initialized to $[N=1,\epsilon=0.1]$ and the number of Fresnel zones $L$ is set to 7 as it was empirically observed that such value leads to a lower depth estimation error. In the case of noise-free RPSF-blurred inputs, the network achieved a RMSE of 0.392 on the test set. The corresponding learned phase mask parameters are $[N=1,\epsilon=0.92]$: the generated RPSF as well as the height map of the mask are shown in Fig.~\ref{fig:learndedrpsf1}. The network learned a single-helix RPSF with high confinement parameter $\epsilon$ meaning that the peak is spread out across a large area.

\begin{figure*}[h!]
\centering
\captionsetup{justification=centering}
\includegraphics[width=0.9\linewidth]{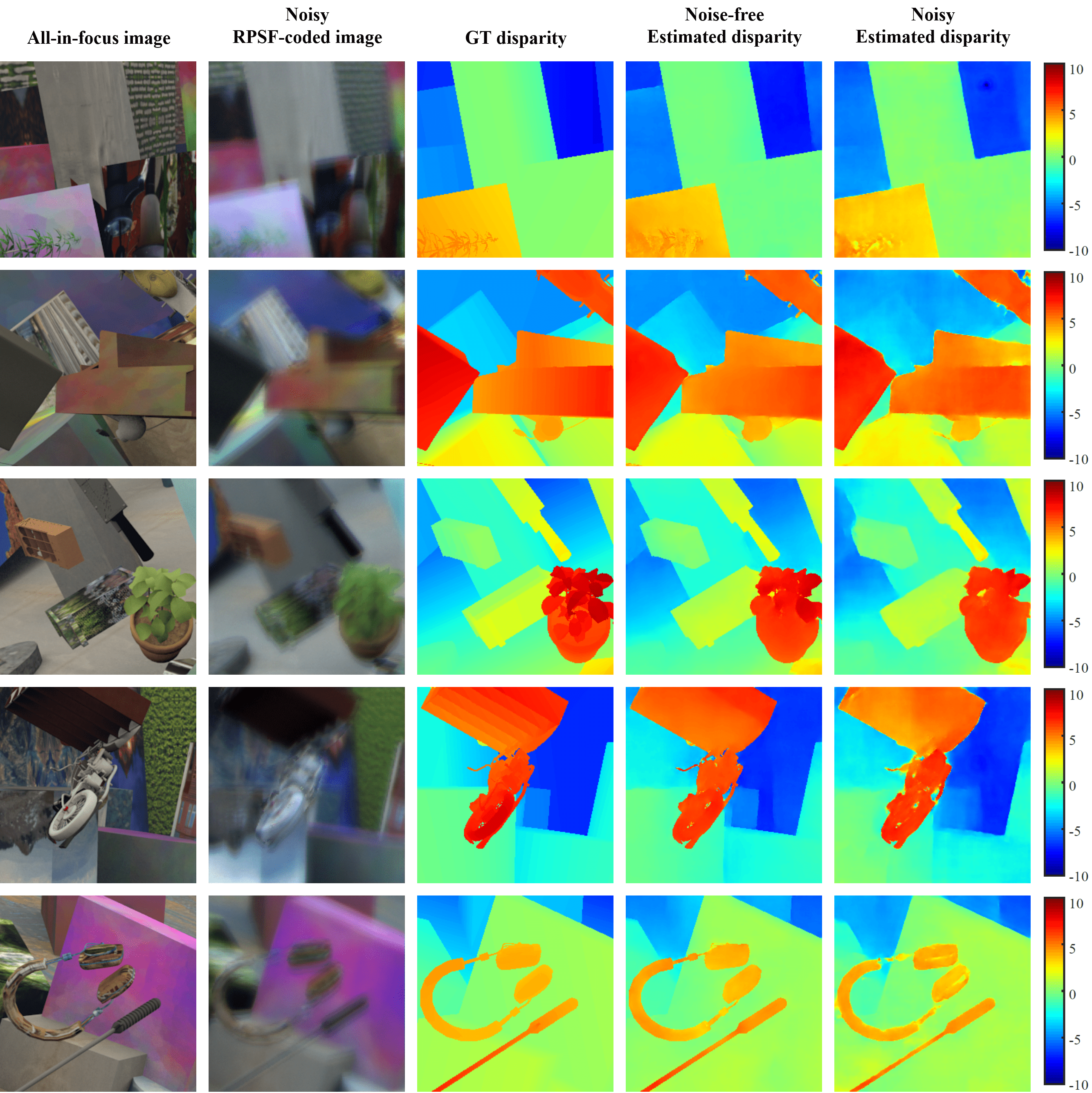}
\caption{Qualitative results on RPSF-blurred images from the test set of FlyingThings3D~\cite{MIFDB16} subset.}
\label{fig:res1}
\end{figure*}

Qualitative results are shown in Fig.~\ref{fig:res1} (additional scenes are shown in the supplementary material): Notice how the network is able to predict accurate disparity maps with fine image details, e.g., the small leaves of the plant shown in the third row or the very fine parts of the headset shown in the last row. However, it becomes harder for the network to predict accurate object boundaries due to the blurring artifacts introduced by the RPSF especially when the blur kernel is larger than the image features. Some artifacts appear at object edges %, the predicted disparity values are sometimes inconsistent with the ground truth, this is
mainly due to the nature of the image formation model used to compose the RPSF-blurred images: the layered depth model used to render RPSF coded images does not accurately simulate the discontinuities around object edges as visible in the RPSF coded images (Fig. \ref{fig:res1}) due to the lack of accurate occlusion modeling. % in the additive approach that we used where sudden blur changes are  visible across object edges at different depth planes. 
Such issue can be addressed by implementing a more advanced and sophisticated blending and matting approaches, e.g. Pyramid-based blending~\cite{kraus2007depth}, at the expense of much higher computation time and complexity for a minor gain in performance.
%, the main drawback of an advanced rendering process is the additional computational time and complexity. Considering that the potential improvement in terms of depth estimation accuracy would be minor, i.e. only a small number of edge pixels would be correctly classified, it makes sense to use rather a faster and simpler rendering techniques that captures the general aspects of OOF imaging such as the one used in our experiments. 

As pointed out in Section~\ref{sec:image_formation}, we also introduced an accurate noise simulation model for  a more realistic evaluation. To handle noise we further fine-tune the DEPTH-DNN on noisy images. In this case we achieve a RMSE of 0.712 on the test set compared to 0.392 achieved on noise-free data. This is of course expected but at the same time the robustness of the system in real world applications should be enhanced.

%\begin{figure*}[h!]
%\centering
%\includegraphics[width=0.8\textwidth]{latex/figures/res_2.png}
%\caption{Qualitative results on noisy RPSF-blurred images compared with noise-free estimated disparity maps. ****TO BE FUSED***}
%\label{fig:res2}
%\end{figure*}

% Need to fuse fig8 and 9 to save space / ok so we only keep 9 since it alreay has 8 in it

In the first two columns of Fig.~\ref{fig:res1}, one can observe image quality degradation after simulating the noisy images where the color down-sampling by the CFA and quantization artifacts by the ADC unit are visible (zoomed-in). %Such noise sources accompanied by the sensor's read and shot noises make it harder for the network to learn accurate disparity values for some small details.
The predicted disparity maps from noisy inputs are shown in the last column. Even though a small performance degradation is noticeable on the noisy predictions, the fine-tuned network is still able to learn fairly accurate disparity maps. %even if at the same time it struggles a bit to predict precise object edges and finer details for the reasons discussed before which are further amplified in this case by the added noise.

%in general we need a more positive/optimistic approach to the discussion of results

%Unlike the simple additive Gaussian noise model used by~\cite{chang2019deep} and \cite{wu2019phasecam3d}, our approach considers multiple noise sources along with the quantization and CFA which leads to more realistic simulations and to a more reliable assessment framework to get an insight on the performance of such model in real world applications:  even though such advanced noise model degrade the quality of the simulated images, the network is still able to produce overall satisfactory results. %and further advanced settings such as image mating and rendering can further enhance the performance of the proposed approach in the task of monocular depth estimation.

%----------------------------
\paragraph{Image restoration}
\label{subsec:image_results}
%----------------------------

\begin{figure*}[h!]
\centering
\includegraphics[width=0.8\linewidth]{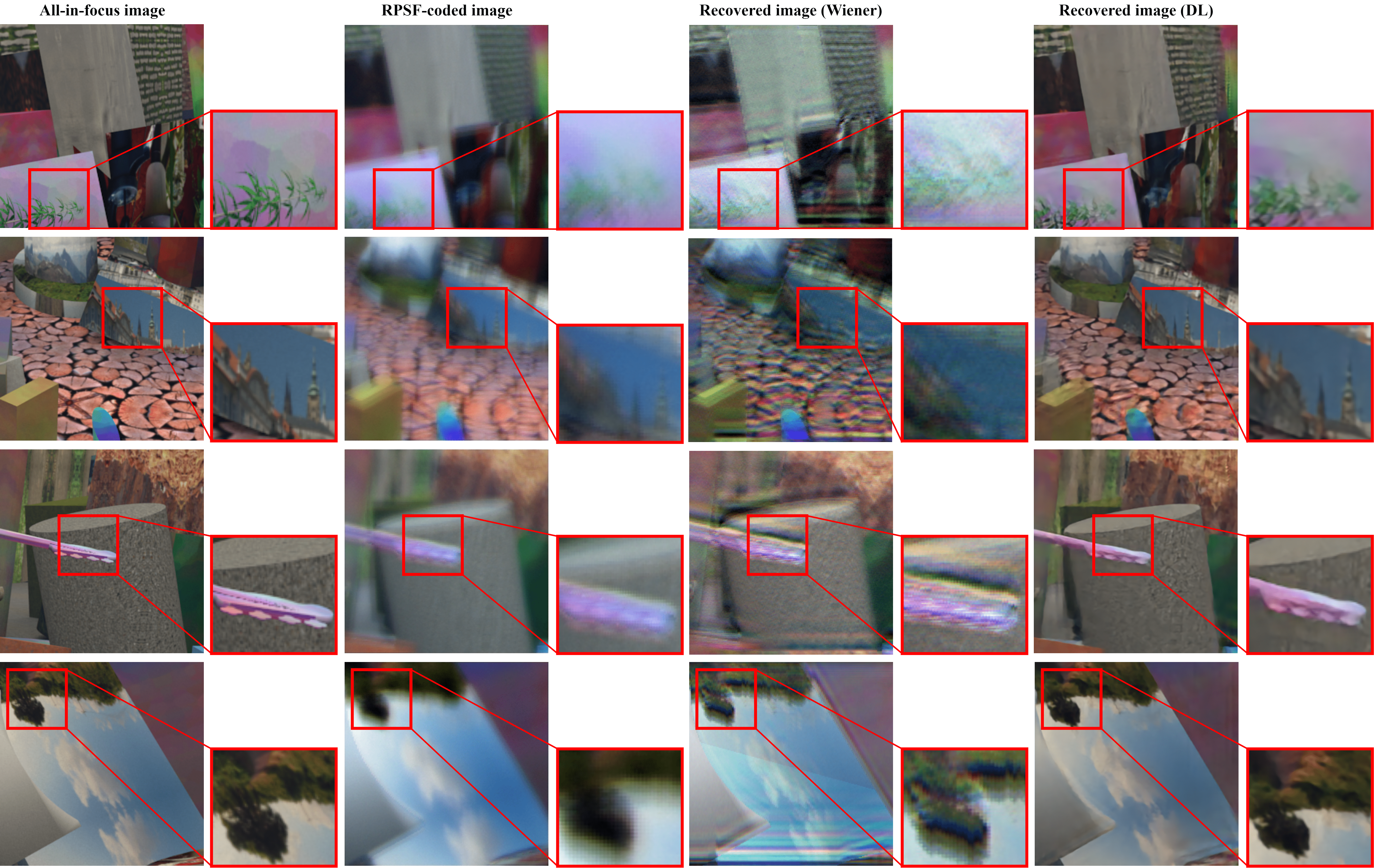}
\caption{Qualitative results of the image deblurring model on the test set of FlyingThings3D~\cite{MIFDB16} subset.}
\label{fig:image_qual}
\end{figure*}

The sharp all-in-focus images are recovered by IMAGE-DNN trained on the subset of FlyingThings3D~\cite{MIFDB16}. Quantitative and qualitative results are shown in Table.~\ref{tab:res6} and Fig.~\ref{fig:image_qual}, respectively. 

The simulated RPSF-coded images have a low mean PSNR of roughly $19$ dB with respect to their sharp noise-free counterparts which were used as the ground truth images during training. %This corresponds to the fact that the network is trained to not only deblur the input images but also to perform noise reduction and to restore the full color resolution. 
As reported in Table.~\ref{tab:res6}, the mean PSNR of the recovered images increased by about $5.5$ dB reaching $24.46$ dB. Also, the Structural Similarity Index Measure (SSIM) achieved is $0.760$ compared  to $0.611$ for the blurred and noisy images. Deblurring results from the traditional Wiener deconvolution filter~\cite{wiener1964extrapolation} are also shown in Fig.~\ref{fig:image_qual}: even if we apply the deconvolution process independently for each depth plane and generated the final result following Eq.~\ref{18}, it results in a low quality image reconstruction with heavy ringing artifacts (see Fig.~\ref{fig:image_qual}).
%to highlight the performance gain that could be reached with new deep learning-based approaches. Since Wiener deconvolution  
This happens since %, even if it tries to make the image sharper, 
the Wiener filter fails to handle the spatially variant blur producing significant ringing artifacts. 

\begin{table}[h!]
\begin{center}
\resizebox{\linewidth}{!}{
\begin{tabular}{|l||c|c|}
\hline
  & PSNR (dB) & SSIM \\
\hline\hline
RPSF-coded images & 19.01 & 0.611 \\
\hline
Recovered images (Wiener) & 18.75 & 0.449 \\
Recovered images (Our approach) & \textbf{24.46} & \textbf{0.760} \\
\hline
\end{tabular}
}
\end{center}
\centering
\caption{Quantitative results of the image deblurring model on the test set of FlyingThings3D~\cite{MIFDB16} subset.}
\label{tab:res6}
\end{table}

Fig.\ref{fig:image_qual} shows some recovered images along with the corresponding blurred inputs and the sharp all-in-focus ground truth. Upon visual inspection, one can notice that the model successfully restores very fine image details and high frequency components, e.g., the small tree leaves shown in the first row, or the various background details present in the second row. Notice also how large regions with smooth as well as textured structures are recovered. %, e.g.,  in the case of the backgrounds of all shown samples. 
The quantization noise and color down-sampling by the CFA make the task even more challenging resulting in some ringing artifacts on object edges.
%had a significant effect on the reconstruction quality as a whole, notice how object edges are sometimes not well defined in the recovered image instead a smoothing-like effect can be seen around fine object details and edges. 
Although an edgetaper~\cite{reeves2005fast} technique was used to limit such artifacts, few are still present in the recovered images, but are significantly reduced when comparing with the ones produced by the Wiener filter. %: those artifacts are mainly present around object edges and image boundaries. 
%Furthermore, it is worth noting that the image deblurring network is much more complex than the depth estimation one and inference time for image restoration is greater than that of depth map estimation.

%-----------------------------------
\subsection{Experimental Results on Real Data}
%-----------------------------------

%-----------------------------------
\paragraph{NYUV2 depth dataset}
%-----------------------------------

In order to compare our approach with state-of-the-art methods, the DEPTH-DNN is trained on a subset of NYUV2 indoor dataset~\cite{eigen2014depth} in an end-to-end fashion with the phase mask's height map. In this experiment, only 10 depth planes are considered in the layered depth presentation (Eq.~\ref{18}) due to memory constraints. The simulated camera lens parameters are the following: $f/4.0$ with  $4$ mm aperture diameter and $16$ mm focal length focusing at a distance of 5 meters. The RGB images are directly convolved with a RPSF cube of the shape $10\times23\times23\times3$. 
%Kinect V1 sensor produces invalid pixels around object edges, reflective surfaces, and distances exceeding the maximum attainable depth by the sensor, this is due to depth uncertainties around such regions. 
In the following evaluations,  we applied the same crop used in competing works (e.g. ~\cite{eigen2015predicting}) % of $[45, 471,  41, 601]$ is used
%to trim invalid border regions 
and excluded the invalid pixels from the Kinect V1 sensor %used in this dataset to build the ground truth %are omitted 
%when calculating the performance metrics
as done by all competing approaches. %Also, in both the ground truth and estimated depth maps. % when evaluating the model performance on the Eigen test set~\cite{eigen2015predicting}. 

\begin{figure}[h!]
\centering
\includegraphics[width=\linewidth]{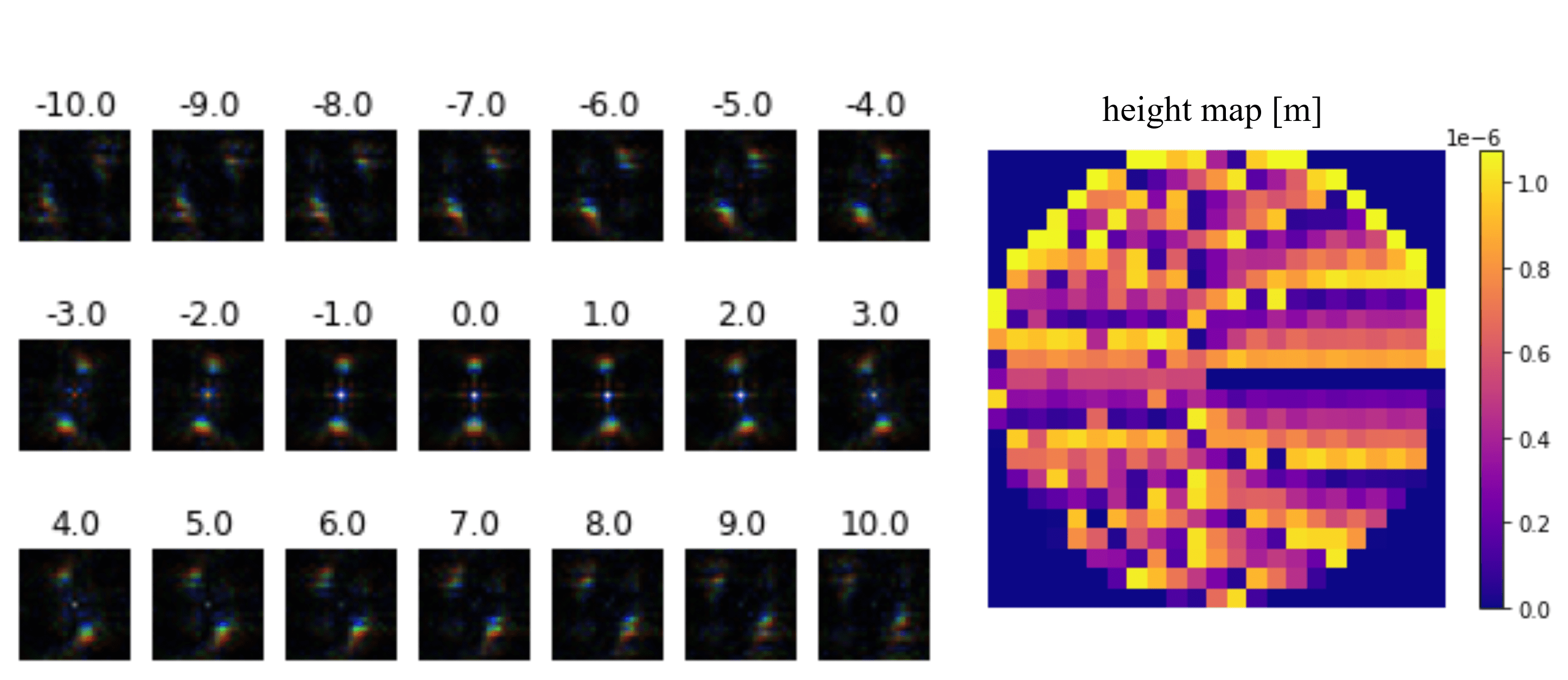}
\caption{Learned RPSF shape at different defocus planes (left), the height map of the phase mask (right).}
\label{fig:learndedrpsf2}
\end{figure}

Fig.~\ref{fig:learndedrpsf2} shows the learned RPSF and the corresponding phase mask. For this dataset, the learned RPSF is a double-helix $[N=2,\epsilon=0.99]$ with two main side lobes that rotate with defocus. We argue that such behaviour is mainly related to the characteristics of the training data: for more complex depth scenes, like in this case, the network tends to converge to higher number of peaks as it makes it easier to correlate between the rotation angle and the corresponding depth plane. Similar to the previous  scenario, the network also converges towards a high confinement parameter $\epsilon=0.99$ producing more spread out peak regions.   

Quantitative results on NYUV2~\cite{eigen2015predicting} test set are reported in Table.\ref{tab:res4}. 
Coded-aperture based competing approaches~\cite{chang2019deep,wu2019phasecam3d} optimized a free-form phase mask parameterized with a superposition of a set of Zernike polynomials~\cite{iskander2001optimal,born2013principles} using a  U-Net~\cite{ronneberger2015u} architecture. Besides using a more accurate camera model, differently from \cite{chang2019deep,wu2019phasecam3d}, our approach learns only a few design parameters for the mask and simultaneously tackles the problem of image quality degradation.
Table.~\ref{tab:res4} shows quantitative results for the different error metrics used in %: the RMSE, the mean relative error (Rel), the mean log10 error (Log10), and the accuracy metrics under thresholds ($\delta_{1}<1.25$, $\delta_{2}<1.25^2$, $\delta_{3}<1.25^3$), refer to~
\cite{eigen2014depth}. % for the corresponding expression for each metric. 
Our approach outperforms the competing methods of~\cite{chang2019deep,wu2019phasecam3d} in all but the last two accuracy metrics (where all top approaches including ours are very close to 1 making them not too significant). 
Note that for the more significant accuracy metric $\delta_{1}$ (i.e., with the lowest threshold value), our approach achieves the highest score, even if we trained the network using a subset of 50k training samples which is less than half of the default split of 120k training samples~\cite{eigen2014depth} used by \cite{chang2019deep} and \cite{wu2019phasecam3d}. In particular, we achieve a significantly lower RMSE value of 0.267 which is down by respectively 0.166 and 0.115 from the ones achieved by~\cite{chang2019deep} and~\cite{wu2019phasecam3d} and is the lowest yet achieved on this challenging dataset for the task of monocular depth estimation.

This performance gain is primarily related to the optimized PSF shape, The one obtained by~\cite{wu2019phasecam3d} has a generic shape with no clear correlations between different defocus planes, while the one obtained by~\cite{chang2019deep} has an elliptical shape with varying section which increases with depth. %, those ellipses carry also depth dependent color information generated by chromatic aberrations at different depth planes. 
On the other hand, the RPSF shape produces a clear and simple correlation between the defocus plane and the corresponding angle of rotation of the main peaks, thus encoding robust depth cues within input images. This, thanks also to the small number of parameters to be learned, explain why the network achieves better depth estimation accuracy with significantly less training data.

\begin{table}[h!]
\begin{center}
\resizebox{\linewidth}{!}{
\begin{tabular}{|r||c|c|c||c|c|c|}
\hline
Method & RMSE~$\downarrow$ & Rel~$\downarrow$ & Log10~$\downarrow$ & $\delta_{1}\uparrow$ & $\delta_{2}\uparrow$ & $\delta_{3}\uparrow$ \\
\hline\hline
Eigen et al.~\cite{eigen2015predicting} & 0.641 & 0.158 & - & 0.769 & 0.950 & 0.988 \\
Laina et al.~\cite{laina2016deeper} & 0.573 & 0.127 & 0.055 & 0.811 & 0.953 & 0.988 \\
Hao et al.~\cite{hao2018detail} & 0.555 & 0.127 & 0.053 & 0.841 & 0.966 & 0.991 \\
DORN~\cite{fu2018deep} & 0.509 & 0.115 & 0.051 & 0.828 & 0.965 & 0.992 \\
Qi et al.~\cite{qi2018geonet} & 0.569 & 0.128 & 0.057 & 0.834 & 0.960 & 0.990 \\
Yin et al.~\cite{yin2019enforcing} & 0.416 & 0.108 & 0.048 & 0.875 & 0.976 & 0.994 \\
BTS~\cite{lee2019big} & 0.392 & 0.110 & 0.047 & 0.885 & 0.978 & 0.994 \\
DAV~\cite{huynh2020guiding} & 0.412 & 0.108 & - & 0.882 & 0.980 & 0.996 \\
Alhashim et al.~\cite{alhashim2018high} & 0.382 & 0.093 & 0.050 & \underline{0.932} & \underline{0.989} & 0.997 \\
AdaBins~\cite{bhat2020adabins} & 0.364 & 0.103 & \underline{0.044} & 0.903 & 0.984 & 0.997 \\
DPT-Hybrid~\cite{ranftl2021vision} & \underline{0.357} & 0.110 & 0.045 & 0.904 & 0.988 & \underline{0.998} \\
\hline
DeepOptics~\cite{chang2019deep} & 0.433 & \underline{0.087} & 0.052 & 0.930 & \textbf{0.990} & \textbf{0.999} \\
PhaseCam3D~\cite{wu2019phasecam3d} & 0.382 & 0.093 & 0.050 & \underline{0.932} & \underline{0.989} & 0.997 \\
\hline
Ours & \textbf{0.267} & \textbf{0.072} & \textbf{0.029} & \textbf{0.952} & \underline{0.989} & 0.997 \\
\hline
\end{tabular}
}
\end{center}
\centering
\caption{Quantitative comparison with the state-of-the-art for monocular depth estimation task on NYUV2  test set~\cite{eigen2015predicting}.}
\label{tab:res4}
\end{table}

\begin{figure*}[h!]
\centering
\includegraphics[width=0.9\linewidth]{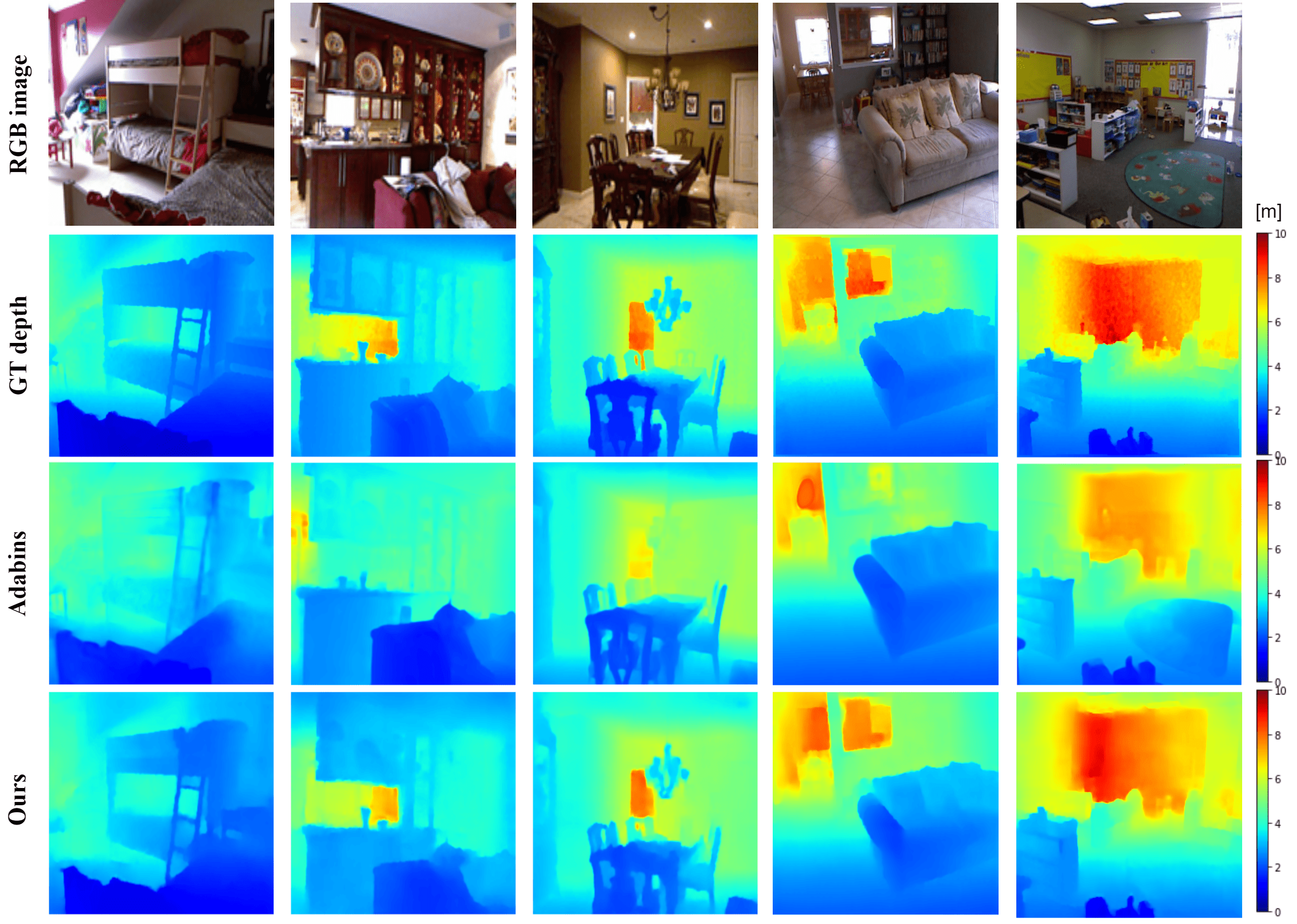}
\caption{Qualitative results from the proposed approach as well as those from AdaBins~\cite{bhat2020adabins} on the test set of NYUV2~\cite{eigen2015predicting}.}
\label{fig:nuy_qual}
\end{figure*}

Table.~\ref{tab:res4} also presents  quantitative results from the state-of-the-art that mainly used sharp all-in-focus images as input. Our approach outperforms all competing methods in all evaluation metrics except for $\delta_{3}$ accuracy metric where all the top approaches including ours are extremely close. More specifically, we achieve substantially lower error metrics compared to the competing approaches which used similar or larger training sets, e.g. Eigen et al.~\cite{eigen2015predicting}, Laina et al.~\cite{laina2016deeper}, DORN~\cite{fu2018deep}, DAV~\cite{huynh2020guiding}, Alhashim et al.~\cite{alhashim2018high},  AdaBins~\cite{bhat2020adabins} and DPT-Hybrid~\cite{ranftl2021vision}, notice that the latter used a pre-trained model on a large combination of different datasets containing 1.4M samples and fine-tuned it on NYUV2 dataset~\cite{eigen2015predicting}. The rest of the competing approaches used less training samples but at the cost of more complex models with pre-trained weights, e.g. Hao et al.~\cite{hao2018detail} used a ResNet~\cite{he2016deep} backbone pre-trained on ImageNet~\cite{deng2009imagenet}.
Our model uses a smaller training set but the core network used for depth estimation is a simple U-Net~\cite{ronneberger2015u} architecture with approximately 8.6M trainable parameters. In contrast, AdaBins~\cite{bhat2020adabins} for example, like most other competing methods used a more complex network architecture with approximately 78M trainable parameters, %to account for the scarcity of depth cues in single input sharp images,
making it slower in both training and inference. In our case, the RPSF encodes strong and robust depth cues making it easier for a simple network to predict accurate depth maps compared to those using conventional RGB inputs. 

A qualitative comparison with AdaBins~\cite{bhat2020adabins} (trained with the same 50k samples as in this approach) is shown in Fig.~\ref{fig:nuy_qual} while further qualitative results are shown in the supplementary materials. Upon visual inspection, one can easily see that our approach produces more accurate and realistic depth maps with respect to the ground truth (the ground truth depth maps are inpainted for visualization purposes, in the supplementary material the ground truth data used for evaluation are shown). 
Due to the scarcity of reliable depth cues in single all-in-focus input images, Adabins~\cite{bhat2020adabins} struggles to predict accurate and sometimes realistic depth values in a consistent manner and fails to predict correct values for images where depth values span large ranges, e.g. the results shown in the two last columns in Fig.~\ref{fig:nuy_qual}. Moreover, sometimes Adabins~\cite{bhat2020adabins} produces erroneous depth predictions where the scene semantics are somehow confusing and the network fails to infer realistic values: such behaviour exposes the main limitation of semantic-based approaches, as visible in the last column of Fig.~\ref{fig:nuy_qual} where the green carpet was misclassified. In contrast, our network consistently produces accurate depth maps for small and large depth ranges alike and is more agnostic to the scene's semantics. %However, it is worth noting that 
The main drawback of our method %, compared to Adabins~\cite{bhat2020adabins}, 
is that it sometimes fails to learn well defined object boundaries due to the blurring artifacts introduced by the RPSF kernel. 
% in the input coded images, whereas the competing approach produces more accurate edges for fine and complex structures in the input image. 

%---------------------------------- 
\paragraph{SUNRGBD dataset}
%----------------------------------
We evaluate the generalization capability of our model for the task of monocular depth estimation where DEPTH-DNN that was previously trained on the subset of NYUV2~\cite{eigen2014depth} is evaluated using the test set of SUNRGBD~\cite{song2015sun} without any further fine-tuning. Quantitative and qualitative results are present in Table.~\ref{tab:res5} and Fig.~\ref{fig:sunrgbd_qual} (additional qualitative results are in the supplementary material). Metric values for competitors shown in Table.~\ref{tab:res5} are taken from~\cite{bhat2020adabins} where methods with publicly available pre-trained models on NYUV2~\cite{eigen2014depth} have been evaluated on the SUNRGBD dataset.

\begin{table}[h!]
\begin{center}
\resizebox{\linewidth}{!}{
\begin{tabular}{|r||c|c|c||c|c|c|}
\hline
Method & RMSE~$\downarrow$ & Rel~$\downarrow$ & Log10~$\downarrow$ & $\delta_{1}\uparrow$ & $\delta_{2}\uparrow$ & $\delta_{3}\uparrow$ \\
\hline\hline
Chen et al.~\cite{chen2019structure} & 0.494 & 0.166 & 0.071 & 0.757 & 0.943 & \underline{0.984} \\
Yin et al.~\cite{yin2019enforcing} & 0.541 & 0.183 & 0.082 & 0.696 & 0.912 & 0.973 \\
BTS~\cite{lee2019big} & 0.515 & 0.172 & 0.075 & 0.740 & 0.933 & 0.980 \\
AdaBins~\cite{bhat2020adabins} & \underline{0.476} & \underline{0.159} & \underline{0.068} & \underline{0.771} & \underline{0.944} & 0.983 \\
\hline
Ours & \textbf{0.335} & \textbf{0.114} & \textbf{0.034} & \textbf{0.937} & \textbf{0.981} & \textbf{0.992} \\
\hline
\end{tabular}
}
\end{center}
\centering
\caption{Quantitative comparison with the state-of-the-art methods for monocular depth estimation task on SUNRGBD~\cite{song2015sun} test set.}
\label{tab:res5}
\end{table}

As shown in Table.~\ref{tab:res5}, our approach outperforms the state-of-the-art in all evaluation metrics with a significant reduction in error metrics, particularly the RMSE (where it achieved $0.335$ compared to $0.476$ of the best competitor) and Log10 ($0.034$ against $0.068$). % as the proposed model scored respectively   and  in those metrics. 
Notice also the accuracy metric $\delta_{1}$ corresponding to the smallest threshold of $1.25$ achieved by our approach which is up by $0.166$ ($\delta_{1} = 0.937$) compared to the best performing approach where the accuracy value achieved is $0.771$, which indicates that ours produces a higher percentage of accurately predicted depth pixel values.
%with respect to the ground truth. 
Such results support the suitability of such engineered PSFs for depth acquisition applications enabling reliable and robust passive monocular depth estimation performance with real-time capabilities. 

\begin{figure*}[h!]
\centering
\includegraphics[width=\textwidth]{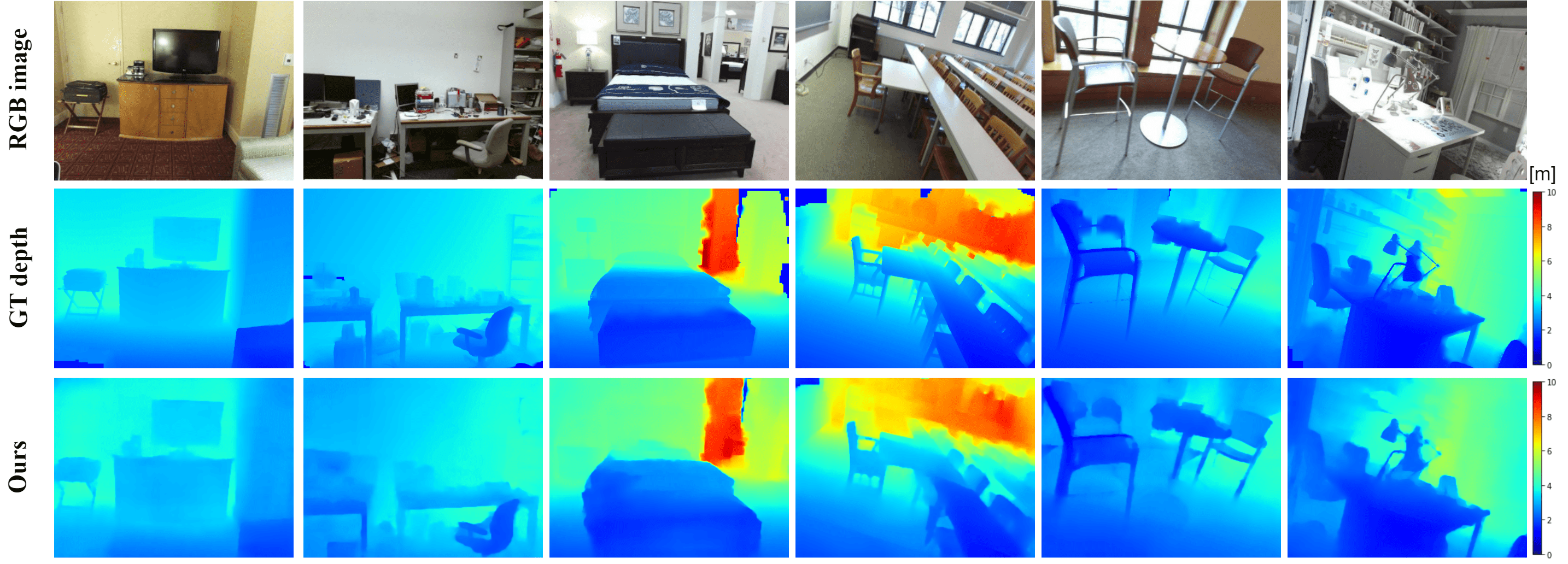}
\caption{Qualitative results from our proposed method on the test set SUNRGBD~\cite{song2015sun}}
\label{fig:sunrgbd_qual}
\end{figure*}
% crop rgb sharp images
Fig.~\ref{fig:sunrgbd_qual} shows some prediction samples from SUNRGBD~\cite{song2015sun} test set, as in the previous case, the network is able to predict overall accurate depth maps but with higher mean RMSE compared to the test set of NYUV2 test set~\cite{eigen2015predicting} which is expected due 
to the different statistical properties between the two datasets.

%in part to the difference in the size of both sets and in an other part to the disparities between the two datasets. 

%mettiamo alla fine nel future work

%Such limitation needs to be addressed in order to enable real-time and high fps applications which is going to be the focus of potential future work.
%------------------

\begin{figure}[h!]
\centering
\includegraphics[width=0.45\textwidth]{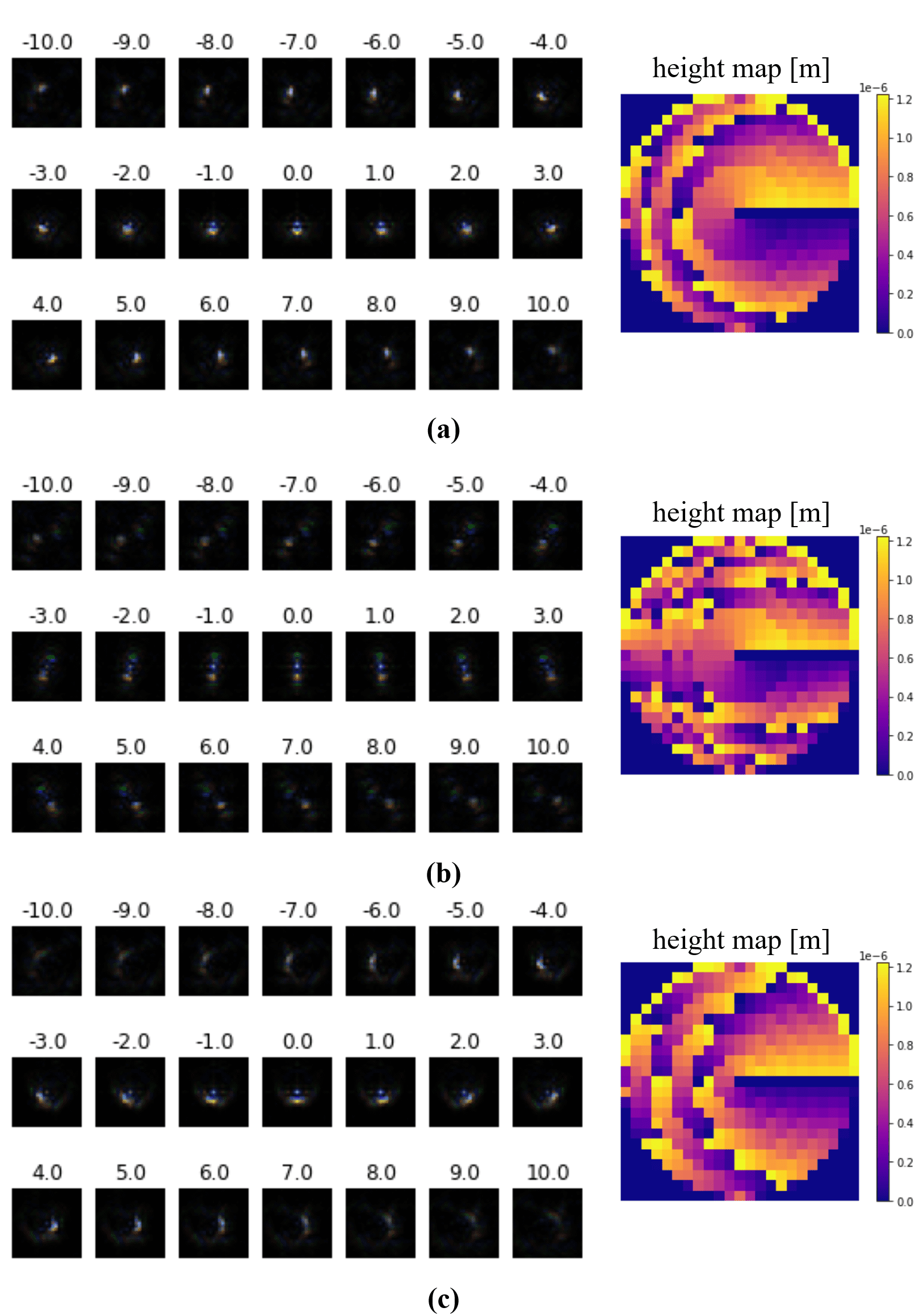}

\caption{The generated RPSFs for the ablation experiments: (a) Single-helix RPSF generated by the fixed mask 1 and its height map. (b) Double-helix RPSF generated by the fixed mask 2 and its height map. (c) Single-helix RPSF generated by the learned mask and its height map.}
\label{fig:ablation_psfs}
\end{figure}

%------------------
\subsection{Ablation Study}
\label{sub_sec:ablation}
%------------------

A number of experiments were carried out as an ablation study to assess the contribution of the various components in our framework.  Quantitative results are shown in Table.~\ref{tab:ablation}, while  the qualitative ones are in the Supplementary Material. In all simulations of the ablation study, the network architecture as well as the training settings and hyper-parameters are the same as indicated in the previous section except for the usage of
% The ablation is typically done with the same params
%\begin{itemize}
  %  \item A
  a simple Gaussian additive noise model  instead of the full  image formation procedure. % . and % as in the full architecture of the proposed solution.
    %\item 
    Furthermore, we set the number of Fresnel zones in the mask design  to $L=5$.
%\end{itemize}

\begin{table}[h!]
\begin{center}
\resizebox{\linewidth}{!}{
\begin{tabular}{|c||l||c|}
\hline
Exp. number & Input & RMSE~$\downarrow$ \\
\hline\hline
1 & All-in-focus & 2.649 \\
2 & Fixed mask 1 $[N=1,L=5,\epsilon=0.5]$ & 1.117 \\
3 & Fixed mask 2 $[N=2,L=5,\epsilon=0.5]$ & 0.815\\
4 & Learned mask $[L=5]$ & \textbf{0.699} \\
\hline
\end{tabular}
}
\end{center}
\caption{Quantitative results of the ablation experiments on the test set of FlyingThings3D~\cite{MIFDB16}.}
\label{tab:ablation}
\end{table}

The baseline is a U-Net trained on all-in-focus sharp images from the subset of FlyingThings3D~\cite{MIFDB16}. The RMSE achieved in this first experiment is 2.649. In the second experiment, a coded aperture with a fixed phase mask design $[N=1,L=5,\epsilon=0.5]$ is used to blur the sharp input images with a depth dependent single-helix RPSF. In fact, this particular mask design is the one which was first introduced by~\cite{prasad2013rotating}. The network trained with the fixed mask achieves a RMSE value of 1.117, i.e. down by 1.532 with respect to the baseline.

In the third experiment, the number of RPSF lobes is increased to $N=2$. Both the number of Fresnel zones $L$ and the confinement parameter $\epsilon$ are the same as in the second experiment. The network trained with such phase mask achieves even better RMSE value of 0.815 compared to the 1.117 achieved by the one in the previous experiment (see Table.~\ref{tab:ablation}). Such performance gain could be due to the more discriminative shape of the double-helix RPSF compared to the single-helix RPSF generated by the previous mask as the rotation can be easily noticed from a depth plane to the other. As shown in Fig.~\ref{fig:ablation_psfs}, the RPSF generated by the second mask has two main side peaks rotating counterclockwise as a function of defocus, the same rotation aspect can be observed in the RPSF shape generated by the first fixed mask (Fig.~\ref{fig:ablation_psfs}a) except that in this case only one main side lobe is present. It suggest that, for a fixed phase mask, a double-helix RPSF conveys more discriminative depth cues than a single-helix one.

In the fourth and last experiment,  the phase mask's trainable parameters $N$ and $\epsilon$ are jointly optimized with the weights of the network, the number of Fresnel-zones is fixed to $L=5$. As expected, the network was able to outperform the baseline as well as the ones trained with fixed masks, reaching a RMSE of 0.699. The learned phase mask parameters are $[N=1,\epsilon=0.91]$ meaning that the RPSF (shown in Fig.~\ref{fig:ablation_psfs}c) has a single side lobe that rotate with defocus. Notice that also the confinement parameter $\epsilon$ is high resulting in a more spread out lobe compared to the one generated by the fixed mask. % in the second experiment. 
It is therefore clear that a joint optimization approach helps the network to effectively learn the correlations between the rotation angle of the PSF and the corresponding depth plane leading to better estimation accuracy.

%------------------
\section{Conclusion}
In this paper we presented  a novel computational camera model where an end-to-end learning framework is proposed for the joint optimization of camera optics and image processing algorithms for the tasks of monocular depth estimation from diffracted rotation. The learned phase mask generates multi-order helix rotating PSFs as a function of defocus, encoding strong depth cues within single 2D images and enabling reliable and accurate depth estimation. Experimental results confirmed the capability of the proposed model to outperform existing methods in the task of monocular depth estimation and to generalize well beyond the training environment. The depth estimation model complexity is significantly reduced compared to the state-of-the-art due to the 3D cues encoded by the RPSF, making it suitable for real-time applications without compromising accuracy. Finally, the sharp all-in-focus images are also recovered through a dedicated non-blind and non-uniform image deblurring module.

Further research will focus on the fabrication of the phase mask via photo-lithography which will be mounted on the back side of the camera's aperture, thus adding depth estimation capabilities to a standard RGB cameras.

\bibliography{mybibfile}

%\newpage
\begin{frontmatter}

\title{\\ \textit{Supplementary Material}}

\abstracttitle{Summary}
\begin{abstract}
In this supplementary material document, we start by analyzing the focusing error in Out Of Focus (OOF) imaging in Section \ref{sec:oof}.  
Then in Section \ref{sec:fp} we briefly present the wave-optics based analysis of light field propagation and  the phase transformation introduced by the lens and the phase mask. 
We also show the effects of the phase mask design parameters on the shape of the resulting RPSF in Section \ref{sec:rpsf}. 

Then, we present some additional qualitative results for the tasks of monocular depth estimation and image deblurring (Sections \ref{sec:res1} and \ref{sec:res2}) that were not possible to fit in the main paper due to space limitations. Finally, a visual example of the results of the various tests made in the ablation study is also presented in Section \ref{sec:abl}.
\end{abstract}

\end{frontmatter}
%------------------
\section*{}
\addcontentsline{toc}{chapter}{Out of Focus Imaging}
\setcounter{section}{0}
\section{Out of Focus Imaging}

\label{sec:oof}
%------------------
Consider a simple imaging system consisting of a thin lens with focal length $f$. The light reflected from an object at distance $z_{o}$ in front of the lens is focused into an image plane at a distance $z_{i}$ behind it, i.e., all rays coming from a point-like object are focused into a point in the image plane that satisfy the well-known thin lens equation $(\frac{1}{z_{o}}+\frac{1}{z_{i}} = \frac{1}{f})$.

Away from the in-focus plane, where the thin lens equation is no longer satisfied, a quadratic phase error at the pupil plane is introduced to model OOF objects. Such error is measured by the defocus value~\cite{goodman2005introduction}:

\begin{equation}
    \Psi= \frac{\pi R^2}{\lambda}(\frac{1}{z_{o}}+\frac{1}{z_{i}} - \frac{1}{f})
\end{equation}

Where $R$ is the pupil radius and $\lambda$ is the wavelength of incident light waves. $\Psi$ indicates the severity of the focusing error and it increases in absolute value as objects move away from the in-focus plane.

%------------------
\section{Field Propagation and the Point Spread Function}
\label{sec:fp}
\begin{figure}[htp]
\centering
\includegraphics[width=0.8\columnwidth]{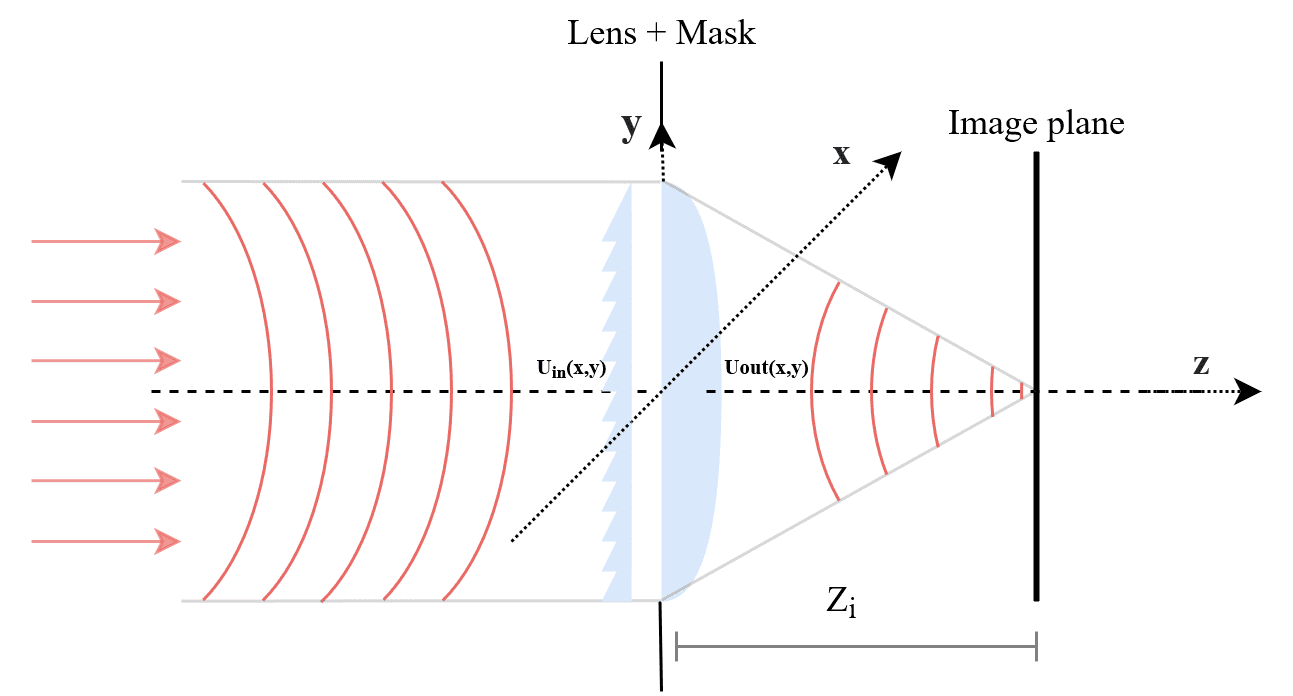}
\caption{Light emitted by an object propagates through free space and its phase is delayed when it passes through the camera optics.}
\label{fig:fresnel}
\end{figure}

%------------------  
In this section, we discuss how a Fresnel diffraction model can be used to simulate near-field wave propagation.
Consider a  thin lens with a thickness profile $h$ situated at the pupil plane of an imaging system (as depicted in Fig.~\ref{fig:fresnel}), notice that $h$ is a function of the  spatial coordinates $(x,y)$, and let $h_{0}$ be the thickest section of the lens. Generally the lens is made of glass with a refractive index $n=1.5$. For a convex thin lens with focal length $f$, $h(x,y)$ is  defined as:

\begin{equation}
  h(x,y)=h_{0}-\frac{x^2+y^2}{2f(n-1)}
\end{equation}

Light wave fields passing through the lens are delayed by an amount proportional to the thickness of the lens $h$ at each point. Thus, a thin lens applies a phase shift to the incident wavefront given by:

\begin{equation}
\label{1.11}
  \Phi_{lens}(x,y)=\frac{2\pi (n-1)}{\lambda}h(x,y)  
\end{equation}

Notice that the phase shift is wavelength-dependent hence using the same lens with polychromatic light results in chromatic aberrations.

By defining the phase shift at the pupil plane, the generalized pupil function, which is a complex function, can be written as:

\begin{equation}
    P(x,y)= A(x,y)e^{i\Phi_{lens}(x,y)}
\end{equation}

Where $A(x,y)$ is the circular aperture mask simulating a finite aperture area. 

In the case of OOF imaging, the pupil function would also introduce a phase error expressed as a quadratic phase term in the generalized pupil function showing the effects of defocus aberrations in the captured image. The new generalized pupil function would have the following form:

\begin{equation}
    P(x,y)= A(x,y)e^{i(\Phi_{lens}(x,y)+\Psi\frac{(x^2+y^2)}{R^2})}
\end{equation}

Where $R$ is the pupil radius and $\Psi$ is the defocus parameter.
In the case of a wave field generated by an ideal point source object, the PSF of the imaging system is expressed as:

\begin{equation}
    PSF(x,y) \propto \lvert \mathscr{F}(P(x,y)) \rvert^2
\end{equation}

Where $\mathscr{F}$ denotes the Fourier Transform.

Concerning the wave field,  let the complex wave field $U_{in}$ propagating in free space %be $U$ and the same field $U_{in}$ is defined 
just before the entrance pupil be:

\begin{equation}
    U_{in}(x,y)= C(x,y)e^{i\Phi_{in}(x,y)}
\end{equation}

Where $C(x,y)$ is the field's amplitude and $\Phi_{in}$ is its phase just before entering the pupil.

In the presence of a phase mask with a phase delay $\Phi_{mask}$ (as shown in Fig.~\ref{fig:fresnel}), the phase delay introduced by the combination of both the lens and mask is expressed as:

\begin{equation}
    \Phi_{optics}(x,y)=\Phi_{lens}(x,y)+\Phi_{mask}(x,y)
\end{equation}

Thus, the expression of the wave field just after the lens system $U_{out}$ in this case is:

\begin{equation}
\begin{split}
    U_{out}(x,y) & =P(x,y)U_{in}(x,y)\\
    & = A(x,y)C(x,y)e^{i(\Phi_{in}(x,y)+\Phi_{optics}(x,y))}
    \end{split} 
\end{equation}

Using Fresnel propagation, $U_{out}$ can be further propagated of a distance $z_{i}$ until it reaches the image sensor (we use a coordinate system $(u,v)$ for the sensor). 

\begin{equation}
\begin{aligned}[b]
    & U_{sensor}(u,v)= \frac{e^{ikz_{i}}}{i\lambda z_{i}}\int_{-\infty}^{+\infty} \int_{-\infty}^{+\infty}U_{out}(x,y)\times {}\\
    & e^{\frac{ik}{2z_{i}}[(u-x)^2+(v-y)^2]}\,dx \,dy
    \end{aligned}
\end{equation}

Where $k=\frac{2\pi}{\lambda}$ is the wave number.

%------------------
\section{Visual Evaluation of the Impact of RPSF Design Parameters}
\label{sec:rpsf}
%------------------

In this section we show some visual examples of  RPSF shapes obtained by using different values of the design parameters.

\begin{figure}[h!]
 \centering
 \includegraphics[width=1\columnwidth]{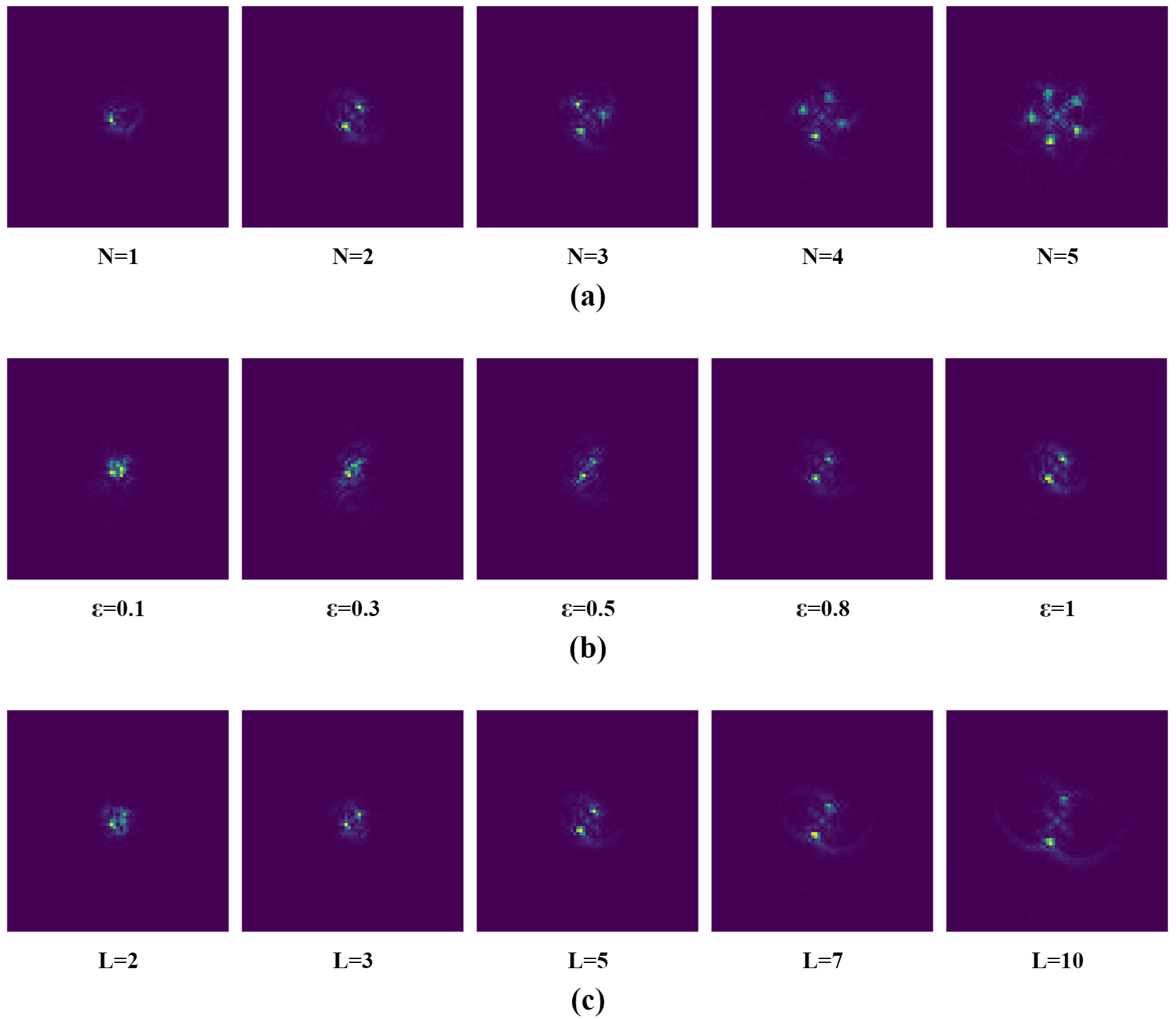}
 \caption{The effects of $[N,L,\epsilon]$  design parameters on the shape of the RPSF: (first row) only the value of $N$ is changed, i.e., $[N=1,2,3,4,5;L=5;\epsilon=0.9]$, (second row) only the value of $\epsilon$ is changed, i.e., $[N=2; L=5; \epsilon = 0.1,0.3,0.5,0.8,1]$, (third row) only the value of $L$ is changed with $[N=2; L=2,3,5,7,10; \epsilon=0.9]$. }
 \label{fig:N_eps_L}
\end{figure}

Recall that we denoted with $N$  the number of rotating peaks within the RPSF:
in the first row of Fig.~\ref{fig:N_eps_L} we show the resulting RPSF shape for different values of $N$.

The design parameter $\epsilon$ controls the peak separation as well as the confinement of each peak as illustrated in the second row of Fig.~\ref{fig:N_eps_L}. It was empirically hypothesized by~\cite{berlich2018high} that the depth range increases as $\epsilon$ decreases.

The number of Fresnel zones $L$ within the phase mask also controls the peak separation as illustrated in the last row of Fig.~\ref{fig:N_eps_L}.
Additionally, the rate of rotation which is $\frac{1}{L}$ in the basic case of a single helix RPSF, can be easily extended to $\frac{1}{NL}$ in the more general case of multiple peaks.
This indicates that the rate of rotation would decrease with higher $L$ and/or $N$ values leading to an increase in the practical depth range with no peak rotation ambiguity. 

In summary, $[N,L,\epsilon]$ can be jointly optimized depending on the target task: they all influence the shape of the RPSF and the practical depth range in which the PSF can be used to encode unambiguous depth information within the captured 2D images.

%------------------
\section{Additional Visual Results for Image Deblurring}
\label{sec:res1}
%------------------

%------------------
%\subsubsection{}
%------------------

\begin{figure*}[h!]
\centering
\includegraphics[width=0.8\linewidth]{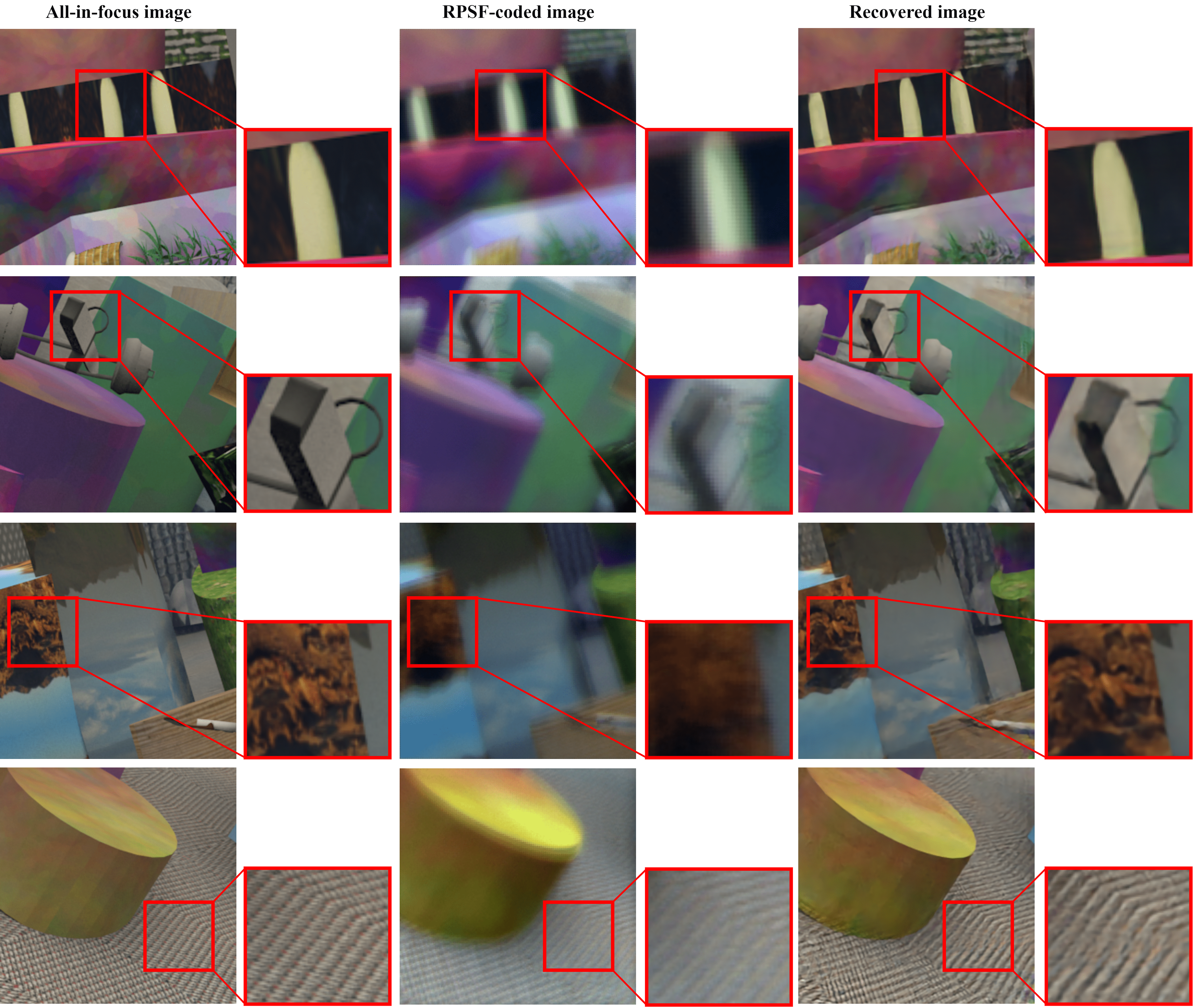}
\caption{Additional qualitative results of the image deblurring model on the test set of FlyingThings3D~\cite{MIFDB16} subset.}
\label{fig:add_deblur_supp}
\end{figure*}

Fig. \ref{fig:add_deblur_supp} shows some additional examples of the proposed image deblurring algorithm on   the FlyingThings3D subset~\cite{MIFDB16}: the network achieves overall satisfactory reconstruction with minor artifacts (a bit of ringing and a few texture patterns not properly recovered are visible upon closer inspection). Some challenging complex structures are restored (e.g. the sample shown in the third row) while other fine details, such as the periodic pattern of the ``floor'' shown in the sample of the last row, proved difficult  to be accurately recovered.

%----------------------------------

\begin{figure*}[h!]
\centering
\captionsetup{justification=centering}
\includegraphics[width=\textwidth]{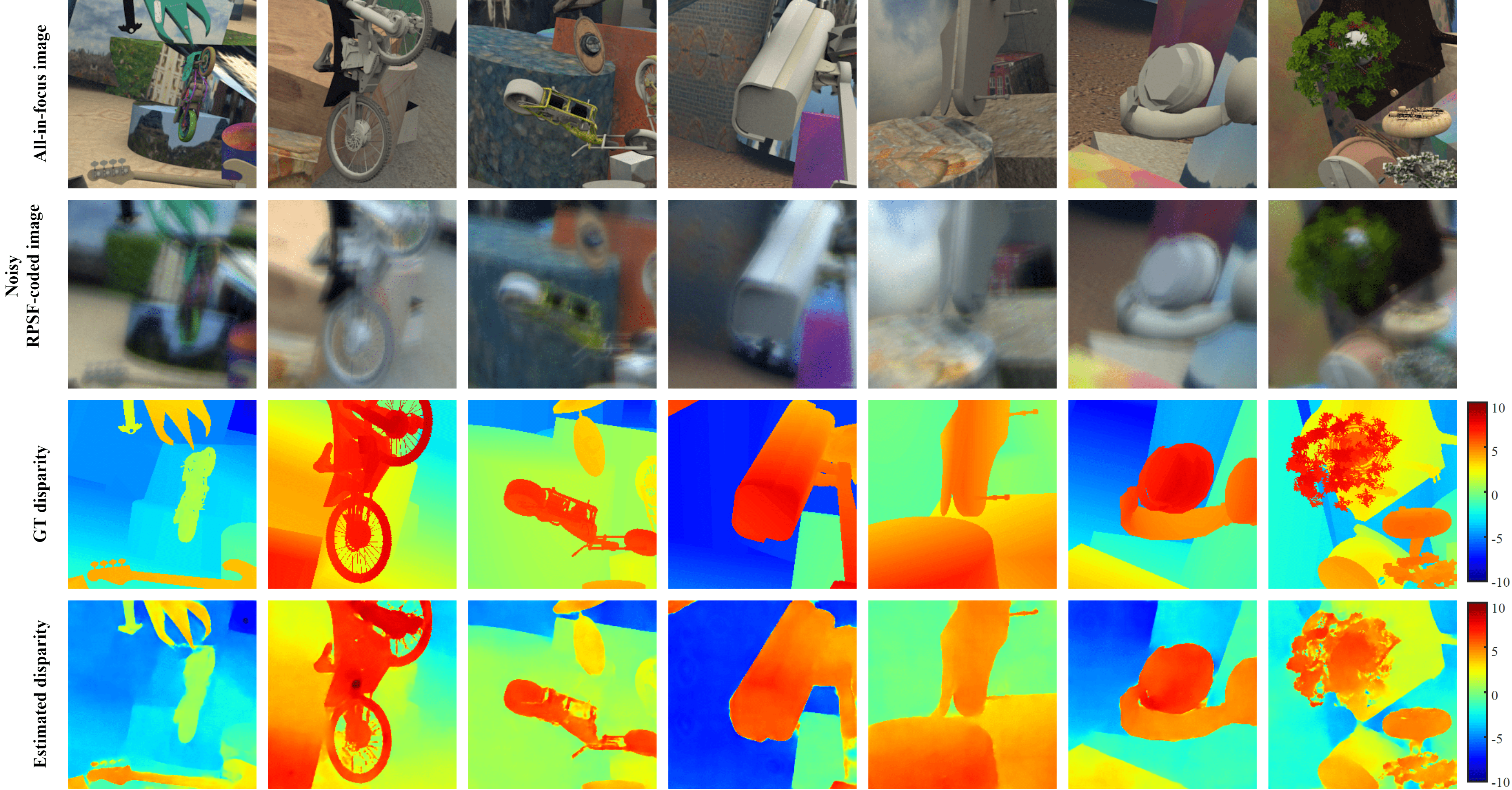}
\caption{Additional qualitative results on RPSF-blurred images from the test set of FlyingThings3D~\cite{MIFDB16} subset.}
\label{fig:supp_ft}
\end{figure*}

%------------------
\section{Additional Visual Results for Monocular Depth Estimation}
\label{sec:res2}
%------------------

Fig. \ref{fig:supp_ft} shows some additional examples of the proposed depth estimation algorithm on  sample images from the  FlyingThings3D test set~\cite{MIFDB16}. These results correspond to the simulated noisy input images where sensor noise, CFA, and quantization noise are added. The network was able to produce accurate disparity maps even for objects with very fine details (e.g. the motorcycles in the second and third columns in addition to the plant in the last column). On the other side, edge pixels are not always accurately estimated (e.g., see the fourth column) as discussed in Section VI in the paper.

\begin{figure*}[h!]
\centering
\includegraphics[width=\textwidth]{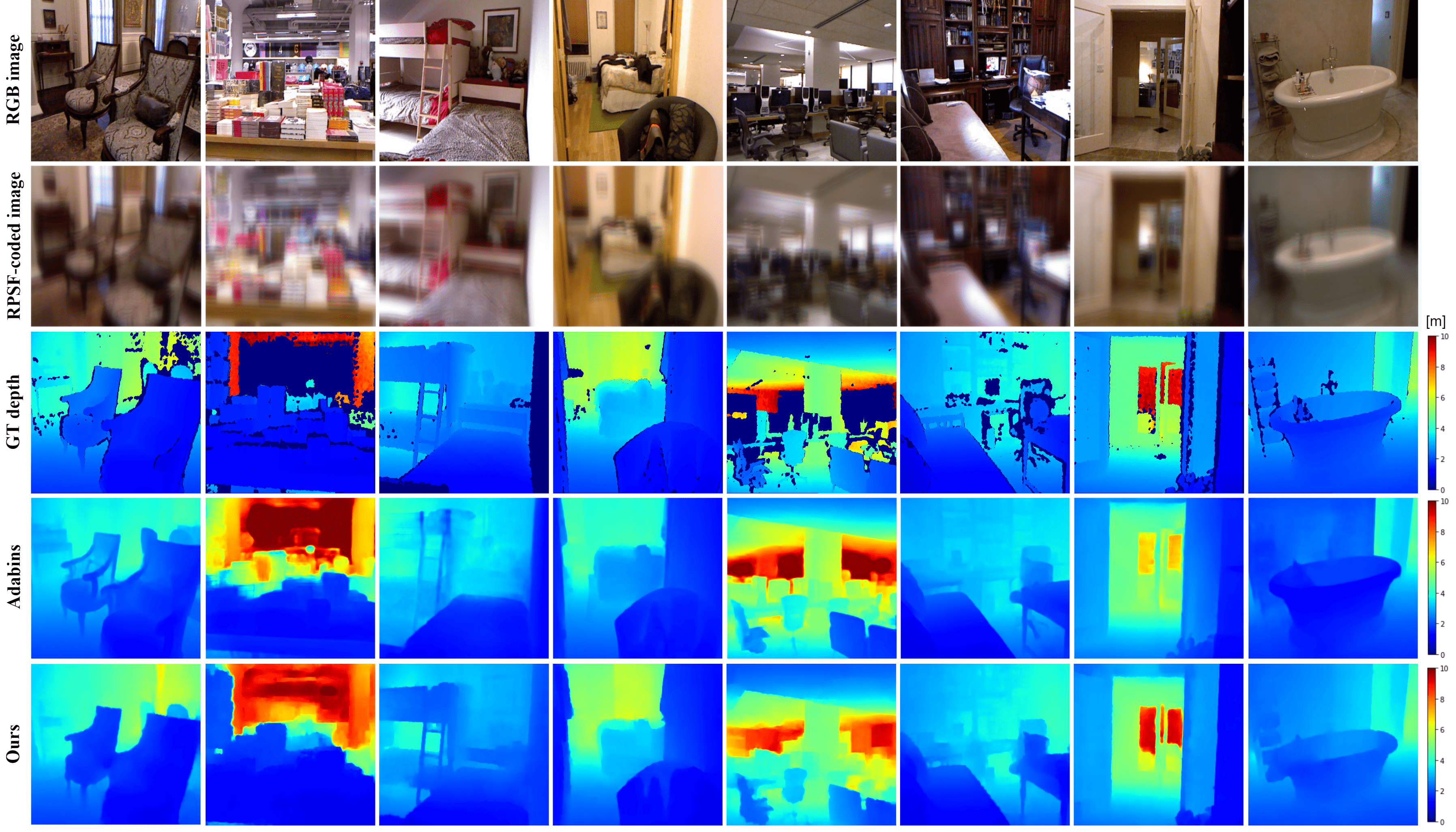}
\caption{Additional qualitative results from the proposed approach as well as those from AdaBins~\cite{bhat2020adabins} on the test set of NYUV2~\cite{eigen2015predicting}.}
\label{fig:supp_nyu}
\end{figure*}

Then, Fig. \ref{fig:supp_nyu} shows some additional examples of the proposed depth estimation algorithm on  sample images from the NYUV2 test set~\cite{eigen2015predicting} and compares them with those from AdaBins~\cite{bhat2020adabins} competing approach. Our approach proved capable of producing superior results for small and large depths ranges alike with no effects or ambiguities whatsoever originating from the scene's semantics.  

\begin{figure*}[h!]
\centering
\includegraphics[width=\textwidth]{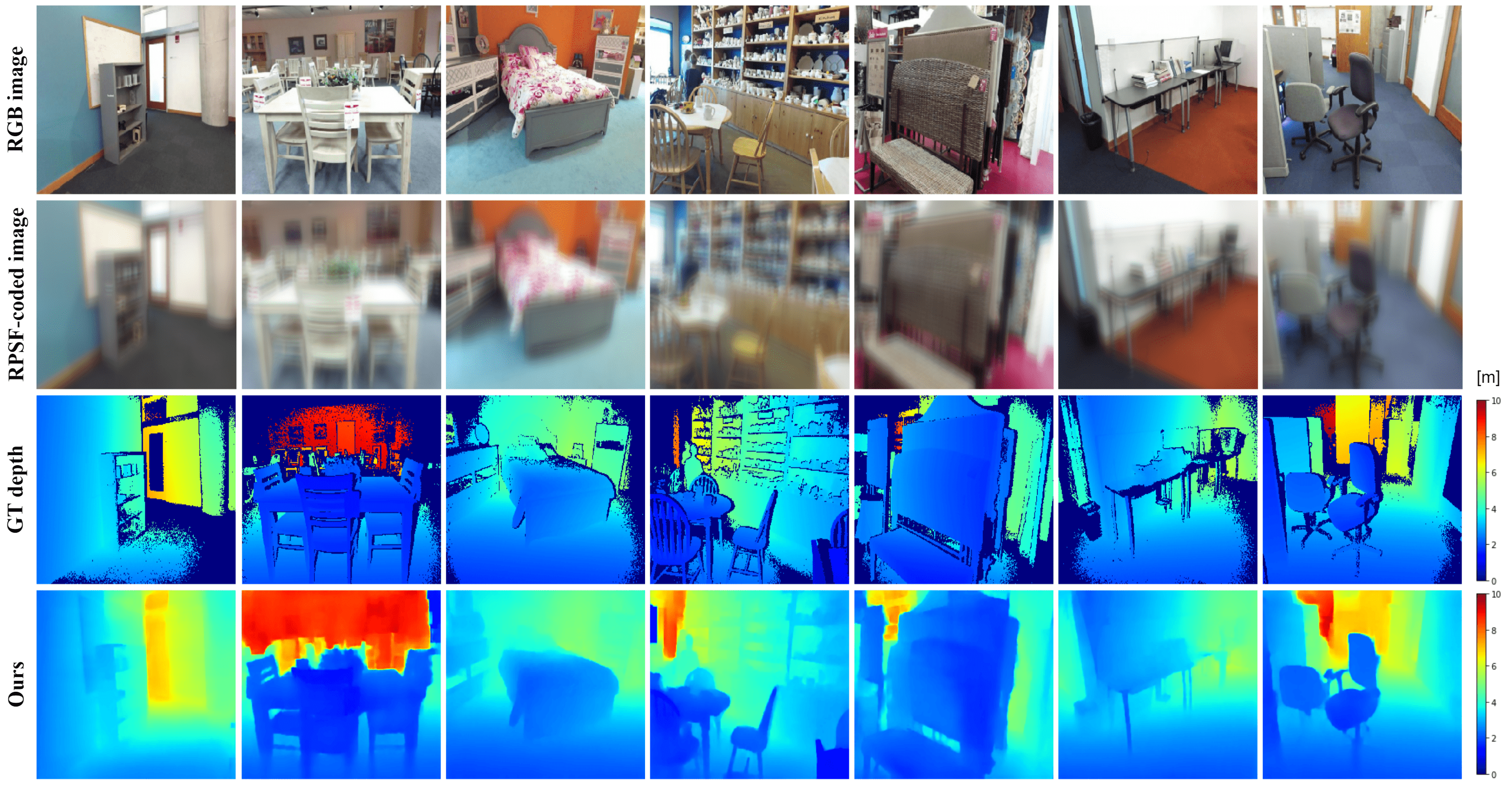}
\caption{Additional qualitative results from our proposed method on the test set of SUNRGBD~\cite{song2015sun}}
\label{fig:supp_sunrgbd}
\end{figure*}

Finally, Fig. \ref{fig:supp_sunrgbd} shows some additional examples of the proposed depth estimation  algorithm on sample images from the SUNRGBD test set~\cite{song2015sun}. Even though the network was not fine-tuned for this specific dataset, it performed very well on this larger  and challenging dataset.

%------------------
%\subsubsection{Monocular Depth Estimation on NYUV2}
%------------------

%------------------
%\subsubsection{Monocular Depth Estimation on SUNRGBD}
%------------------

\section{Qualitative Results for the Ablation Study}
\label{sec:abl}

\begin{figure*}[h!]
\begin{center}
\includegraphics[width=\columnwidth]{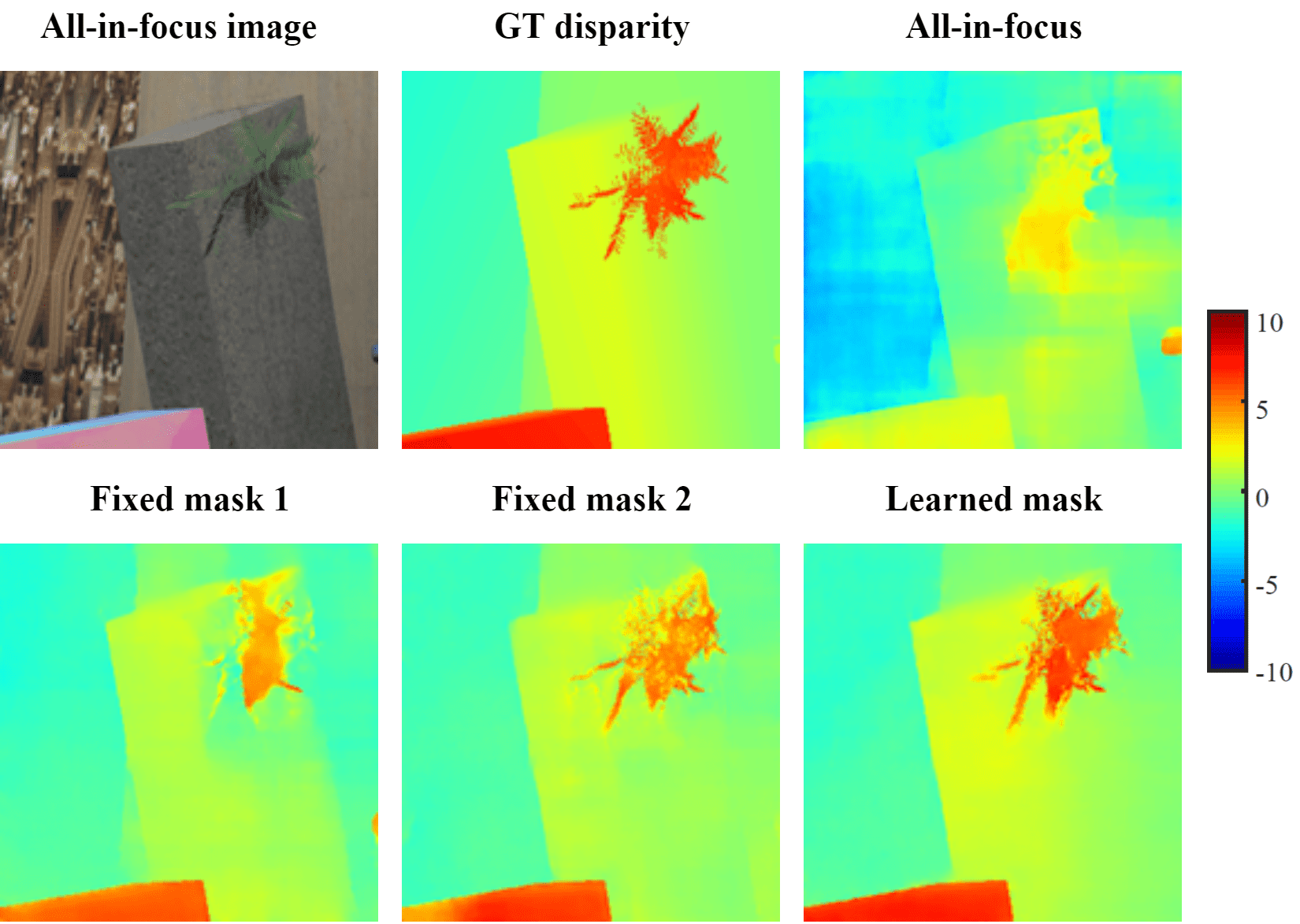}
\end{center}
   \caption{A sample from the qualitative results for the four conducted ablation experiments on the test set of the subset of FlyingThings3D~\cite{MIFDB16}.}
\label{fig:ablation}
\end{figure*}

Fig.~\ref{fig:ablation} shows a sample of the predicted depth maps for the four conducted ablation experiments discussed in the paper. Quantitatively, the network trained using all-in-focus images as input fails to discriminate between positive and negative defocus values, i.e., it is very difficult for the network to accurately predict whether objects are close or far from the camera because of the  lack of reliable depth cues within the clear aperture PSF. While the rest of the performed experiments significantly improve the depth estimation accuracy, the learned mask proved capable not only to predict accurate depth maps but also to preserve  fine details around object edges at different depth planes as shown, for example, in Fig.~\ref{fig:ablation} around the plant edges.

\end{document}